\documentclass[twocolumn]{aastex7}
\usepackage{bm}
\usepackage{ae,aecompl,amsmath}

\begin{document}

\title{The Arc in the DX Cha Circumbinary System: Evidence For a Retrograde Circumbinary Disk}

\author[0000-0002-4489-3491]{Cheng Chen}
\affiliation{School of Physics and Astronomy, University of Leeds, Sir William Henry Bragg Building, Woodhouse Ln., Leeds LS2 9JT, UK}
\affiliation{Department of Physics and Astronomy, Bishop's University 2600 Rue College, Sherbrooke, QC J1M 1Z7, Canada}
\email[show]{phycc@leeds.ac.uk}  
\email[show]{cchen@ubishops.ca}  

\author[0000-0002-0756-9836]{Daniela Paz Iglesias}
\affiliation{School of Physics and Astronomy, University of Leeds, Sir William Henry Bragg Building, Woodhouse Ln., Leeds LS2 9JT, UK}
\email{D.P.Iglesias@leeds.ac.uk}

\author[0000-0002-1575-680X]{James M. Miley}
\affiliation{Departamento de Física, Universidad de Santiago de Chile, Avenida Victor Jara 3659, Santiago, Chile}
\affiliation{Millennium Nucleus on Young Exoplanets and their Moons (YEMS), Chile}
\affiliation{Center for Interdisciplinary Research in Astrophysics and Space Exploration (CIRAS), Universidad de Santiago, Chile}
\affiliation{Joint ALMA Observatory, Alonso de Córdova 3107, Vitacura, Santiago, Chile}
\affiliation{European Southern Observatory, Alonso de Córdova 3107, Vitacura, Santiago, Chile}
\email{james.miley@alma.cl}

\author[0000-0002-2137-4146]{C.J. Nixon}
\affiliation{School of Physics and Astronomy, University of Leeds, Sir William Henry Bragg Building, Woodhouse Ln., Leeds LS2 9JT, UK}
\email{C.J.Nixon@leeds.ac.uk}  

\begin{abstract}
Observations of the binary system DX Cha (HD 104237) reveal a compact, asymmetric ring structure with a radius of 0.43\,au. This ring is just outside the binary orbit, which has semi-major axis $a_{\rm b} = 0.22$\,au and eccentricity $e_{\rm b} = 0.665$; placing the ring at $\approx 1.2$ times the binary apocenter distance. The inner regions of circumbinary disks, $\approx 2-3\,a_{\rm b}$, are typically evacuated by strong gravitational torques from the binary, resulting in a deep gap between the binary and the disk. Accordingly, previous numerical simulations of DX Cha have found an eccentric inner cavity with almost no material inside $\approx 1$\,au, and we find similar results when making the same assumption that the circumbinary disk orbits in the same direction as the binary. However, the disk can exist much closer to the binary if it is retrograde. For DX Cha we find that the inner edge of a retrograde disk occurs at $\approx 2a_{\rm b}$, and moreover takes the form of one or two arcs, in agreement with observations. We therefore suggest that the circumbinary disk in the DX Cha system could be orbiting retrograde to the binary star system in the center. We conclude that compact circumbinary disks  observed in young stellar systems are important targets for future observations; if the disks  are prograde then their properties are likely to be significantly different from current estimates, while if they are retrograde then this will have profound implications for our understanding of star and planet formation.
\end{abstract}


\section{Introduction}
The interaction between a binary system and an external accretion flow is a complex process \citep[e.g.][]{Lin:1986,Pringle1991,Artymowicz1994}. The disk flow tries to move inwards towards the binary through the action of a viscosity \citep{Pringle:1972,LP1974,SS1973}, which is typically mediated by magnetic fields resulting in hydromagnetic turbulence \citep{SS1973,BH1991}. The binary, however, attempts to arrest the disk inflow by transferring energy and angular momentum to the disk orbits through orbital resonances \citep[cf.][]{Lynden-Bell1972,Papaloizou:1977,Goldreich:1979}. The structure of the inner disk regions depends strongly on the competition between these effects and also on the disk properties which affect, for example, where the transferred energy and angular momentum is deposited in the disk \citep[][see \citealt{Heath2020} for a recent discussion]{LP1993,Korycansky1995,Lubow1998,Bate2002}. For typical disk-binary parameters the inner disk regions are depleted of material, and accretion on to the binary takes place via time-dependent streams, which fuel internal disks  around each star \citep[e.g.][]{Artymowicz1996}.

However, this picture of binary-disk interaction assumes that the binary and disk orbit in the same direction. If instead, the binary rotates in the opposite direction to the disk then the orbital resonances are either drastically reduced in strength for eccentric binaries \citep{Nixon2015,Ivanov:2015} or absent altogether for circular binaries \citep{Nixonetal2011a}. In this case, the disk can viscously inflow towards the binary until the inner disk edge is strongly perturbed from a circular orbit by the passing star, typically the lower-mass secondary star. For circular binaries, the material from the disk inner edge is either captured directly by the secondary star into a circumsecondary disk or gravitationally perturbed towards the primary star ending up in a circumprimary disk \citep{Nixonetal2011a}, while for eccentric binaries it can be dragged in and expelled outwards on the timescale of the binary orbit \citep{Nixon2015}. The formation of a retrograde circumbinary disk can occur through a combination of precession and dissipation of a misaligned disk that is initially closer to retrograde than prograde, i.e. the angle between the disk and binary angular momentum vectors is $90^\circ < \theta < 180^\circ$ \citep{Nixonetal2011b}.\footnote{The criterion is more complex in the case that the disk angular momentum, $J_{\rm d}$, is not small compared to the binary angular momentum, $J_{\rm b}$, in which case the requirement for the disk to settle into a retrograde orbit is $\cos\theta < -J_{\rm d}/2J_{\rm b}$ \citep{Nixonetal2011b}. Additionally, for sufficiently eccentric binaries with disks  that are sufficiently close to $90^\circ$ inclinations, the stable configuration for the disk may be polar in which the disk angular momentum vector aligns to the binary eccentricity vector \citep{Aly2015}.} The difference in dynamics between prograde and retrograde disks  leads to a significant difference in the structure of the disk near the inner disk edge around the binary.

Misaligned disks are likely to form in different astrophysical environments. For example, in the centers of merged galaxies disks may form with essentially random inclinations around a supermassive black hole binary \citep{Nixonetal2011a,Nixonetal2011b,Nixon2013,Roedig2014,Dunhill:2014}. However, in nearby star forming regions the initial conditions for star formation are also highly chaotic and turbulent \citep{Bate2003,McKee2007}, which is conducive to the formation of misaligned circumbinary disks  around young stars \citep{Nixon2013,Bate2018}. Indeed, observations indicate that such misalignments are common  \citep[e.g.,][]{Chiang2004,Winn2004,Kennedy2012,Brinch2016}, and in some cases polar disks have been found \citep{Kennedy2019,Kraus:2020}. Population studies have found that there is a transition in the present-day alignment properties of circumbinary disks  with the orbital period of the binary; for periods less than $\sim 100$\,days almost all circumbinary disks  are coplanar with the binary, whereas for longer period systems a substantial fraction of disks  are strongly misaligned to the binary orbit \citep{Czekala2019}.

Circumbinary disks  can be found around many types of young binary systems, including systems containing Herbig Ae stars. These are a subclass of pre-main sequence stars characterized by their intermediate mass, typically ranging from $1.5-3\,$M$_{\odot}$. They are commonly found in star-forming regions and exhibit emission line features from circumstellar environments, often possessing circumstellar disks  composed of gas and dust, resulting in an infrared excess in their spectral energy distributions \citep{Brittain2023, Nidhi2023}. DX Chamaeleontis (DX Cha, HD 104237) is a pre-main-sequence spectroscopic binary with a short period $T_{\rm b} \simeq$ 19.859 days corresponding to a semi-major axis of $a_{\rm b} = 0.22$\,au \citep{Bohm2004}. This system has been observed and comparatively well-studied for more than 20 years by spectroscopic and infrared interferometric observations \citep[e.g.][]{Leinert2003, Bohm2004, Grady2004, Tatulli2007, Garcia2013, Cowley2013, Bailer-Jones2021, Juhasz2025}. The binary consists of a Herbig Ae star with a mass of $2.2\,$M$_{\odot}$ and a K3 star with a mass of $1.4\,$M$_{\odot}$ \citep{Grady2004, Garcia2013}. The two stars are in an eccentric orbit with $e_{\rm b} = 0.665$ \citep{Bohm2004}.  Prior work constrains the binary orbital orientation on the sky and indicates that the disk is viewed close to face-on \citep{Garcia2013}. However, the relative prograde or retrograde sense between the binary and the disk remains uncertain in the absence of resolved disk kinematics. The system is surrounded by a circumbinary disk of mass $M_{\rm d} = 0.04\,$M$_{\odot}$ estimated by modeling the disk spectral energy distribution \citep[SED;][]{Hales2014}. Due to its near face-on geometry, this system provides an ideal case for detailed studies of the accretion process in young binary systems. Here we are specifically interested in the emission on small scales around the binary system. 

\cite{Garcia2013} report a tidal truncation radius of the circumbinary disk around DX Cha at $\approx 0.44$\,au ($2a_{\rm b}$) from analyzing data from the Very Large Telescope Interferometer
(VLTI)/AMBER in the $K$-band continuum and the Br$\gamma$ line. To understand the disk-binary interaction in DX CHa, \cite{Dunhill2015} performed numerical simulations of a prograde circumbinary disk, finding that the asymmetric cavity around the binary is large, with diameter of $\approx 4$\,au which is significantly larger than suggested by the observations presented by \cite{Garcia2013}. Although the simulations in \cite{Dunhill2015} suggest there could be an accretion stream attached to both stars within the cavity, \cite{Juhasz2025} fit more recent {\it L}-band (3.55 $\mu$m) data from the Multi Aperture Mid-Infrared Spectroscopic Experiment (MATISSE) on VLTI and find a best-fit of an azimuthally asymmetric narrow ring structure (width $\sim$ 0.13 au) with a diameter of 0.86 au. \cite{Juhasz2025} conclude that the data shows material orbiting within a circumbinary disk rather than spiraling on to the stars from within a cavity. However, this is at odds with the circumbinary disk structure found in simulations \citep{Dunhill2015}, and might be more in line with the structures formed from a retrograde circumbinary disk. 

To test this possibility, we present Smoothed Particle Hydrodynamics (SPH) simulations of prograde and retrograde circumbinary disks with parameters appropriate to the DX Cha system to see which can provide a better fit to the structures inferred from the observations. In Section~\ref{num} we describe the model setup of our simulations. We present our simulation results in Section~\ref{sim}. Finally, we present our discussion in Section~\ref{dis}, and conclusions in Section~\ref{con}.

\section{Numerical setup}\label{num}
To investigate the DX Cha system, we employ the SPH code \textsc{phantom} \citep{Price2018} which has been widely used for studying circumbinary systems \citep[starting with][]{Nixon2012}. We adopt the binary system parameters from Table 1 in \cite{Juhasz2025}: the total binary mass $M_{\rm b}$ is composed of the primary mass $M_1 = 2.2,$M$_{\odot}$, and the secondary mass $M_2 = 1.4,$M$_{\odot}$. The binary eccentricity is $e_{\rm b} = 0.665$, and the separation of the binary is $a_{\rm b} = 0.22$\,au. 

We set up an accompanying circumbinary disk coplanar with the binary and with the disk material initially following circular orbits. The total disk mass is initially $M_{\rm d}=10^{-3} M_{\rm b}$. The disk initially extends from $R_{\rm in} = 3a_{\rm b}$ to $R_{\rm out} = 10a_{\rm b}$, and is comprised of $N_{\rm p,ini}=5\times10^6$ particles. This choice of $M_{\rm d}$ serves as a fiducial normalization and is of the same order as the mean disk mass inferred from CO emission modeling \citep{Stapper2024}. However, our simulated disk is truncated at $R_{\rm out}\approx 2.2$\,au, which is much smaller than the CO-traced disk extent of $\approx 40$\,au reported for DX~Cha \citep{Stapper2024}; therefore, the adopted $M_{\rm d}$ in our setup should not be interpreted as the total disk mass on $\sim 40$\,au scales. We use a small outer disk radius to focus the resolution of our simulation on the inner regions where the disk--binary interaction predominantly occurs. In general, the choice of $M_{\rm d}$ does not affect the disk evolution unless it is sufficiently high that the disk self-gravity becomes important (which occurs when $M_{\rm d}/M_{\rm b} \gtrsim H/R$; \citealt{Pringle1981}). We do not include the gas self-gravity in our simulations and thus, while the binary orbit is held fixed, we can rescale the disk mass if required. However, when allowing the binary orbit to evolve, the value of the disk mass does affect the binary orbital evolution. We return to this point in Sec.~\ref{acc}.

We set the outer boundary condition as zero-torque, which means that any particles with $R > R_{\rm out}$ are removed along with their angular momentum. The accretion radii of both stars $R_{\rm acc}$ are initially set to $0.4 a_{\rm b}$ to reduce the computational cost of reaching the disk's quasi-steady state at larger radii. Once this state is reached, we reduce $R_{\rm acc}$ for each star to 0.1 of its effective Roche lobe radius. This is given by \citep{Eggleton1983}
\begin{equation}
    R_{\rm L} = \frac{0.49 q_{\rm b}^{2/3}}{0.69 q_{\rm b}^{2/3}+\ln({1+q_{\rm b}^{1/3}})} a_{\rm b}.   
\end{equation}
Thus, $R_{\rm acc}$ for the primary and the secondary are $0.0389 a_{\rm b}$ and $0.0320 a_{\rm b}$, respectively. With these smaller accretion radii, the gas flows around each star can be resolved.\footnote{Typically the accretion radii used in numerical simulations are significantly larger than the physical size of the objects being simulated. However, in this case, using a stellar mass-radius relation of $R_\star/R_\odot = (M_\star/M_\odot)^{0.8}$, the accretion radii of 0.1 times the Roche lobe is very close to the actual stellar radius. The magnetic field of the stars could redirect the flow at several stellar radii \citep[e.g.][]{Bouvier2007}. We do not account for this effect in our simulations, but we note that the structures we are interested in are on scales outside of the binary orbit and thus unlikely to be significantly affected by the stellar magnetic field.}

We use a locally isothermal equation of state for the gas through a modified form of the prescription suggested by \cite{Farris2014} for circumbinary disks. The sound speed $c_{\rm s}$ depends on the distance from the primary star, $r_1$, and the secondary star, $r_2$, and is given by 
\begin{equation}
c_{\rm s} = h R_{\rm in}^q \frac{\left(M_1/r_1 + M_2/r_2 \right)^{q}}{\left(M_1 + M_2\right)^{q}} \sqrt{\frac{GM_{\rm b}}{R_{\rm in}}}, \\
\end{equation}
where we use a disk sound speed power-law index of $q = 0.25$ and $h$ is the disk angular semi-thickness at $R_{\rm in}$ which we take to be 0.05.

The initial surface density $\Sigma(R)$ is set from the steady-state solution to the 1D disk diffusion equation with mass added to the disk at a rate $\dot{M}_{\rm add}$ to keep it steady at a radius $R_{\rm add}$ = 7$a_{\rm b}$. This is done via the method outlined in the Appendix of \cite{Drewes2021}, but here we set the initial velocity of the injected particles to be the velocity of their nearest neighboring particles, rather than the local Keplerian velocity.

The physical viscosity is modeled by a Navier-Stokes viscosity corresponding to a Shakura-Sunyaev disk viscosity parameter $\alpha_{\rm SS}=0.1$ \citep{SS1973}. For the numerical viscosity, we use a variable linear viscosity coefficient with $\alpha_{\rm min} = 0.01$ and $\alpha_{\rm max} = 1$. The value of the linear coefficient is set via a modified \cite{Cullen2010} switch \citep[see][for details]{Price2018}. For the quadratic numerical viscosity, we take $\beta_{\rm SPH} = 2\alpha_{\rm SPH}$ \citep[e.g.\@][]{Chen2025a}.

Our simulations allow material to accrete on to each of the stars. However, to create a quasi-steady-state disk structure\footnote{Here {\it quasi} refers to the fact that the binary orbit still induces time dependence to the disk structure. However, by averaging over (a small multiple of) the binary orbital period, the disk properties have reached a steady state.}, we initially do not include the back reaction from the disk on to the binary evolution. Once a steady state is reached we allow the back reaction to determine the subsequent evolution of the binary orbit. The timescale on which the disk reaches this state is the viscous timescale $t_{\nu}$ = $R^2$/$\nu$, where $\nu$ = $\alpha_{\rm SS} (H/R)^2 \ R^2 \Omega$. Based on these parameters, the disk is expected to reach a steady state after a time $\sim 10,000T_{\rm b}$, where $T_{\rm b}$ is the binary orbital period.

\section{Simulation results }\label{sim}
First we note that the prograde and retrograde disk simulations reach a quasi-steady-state after $\approx 5,000$ and $\approx 5,500$ binary orbits respectively. This is slightly quicker than expected in part due to the presence of numerical viscosity which slightly augments the physical viscosity included in the simulations. At this stage we are still employing a relatively large accretion radius as described above to reduce computational cost. We therefore change the sink particle accretion radii, to 0.1 of their respective Roche lobe radii, and continue the simulation until a new quasi-steady-state is reached.

We describe the resulting disk structures in the sub-sections below, with the prograde case first and then followed by the retrograde case.

\subsection{Prograde case}
\begin{figure*}
    \centering
    \includegraphics[width=\linewidth]{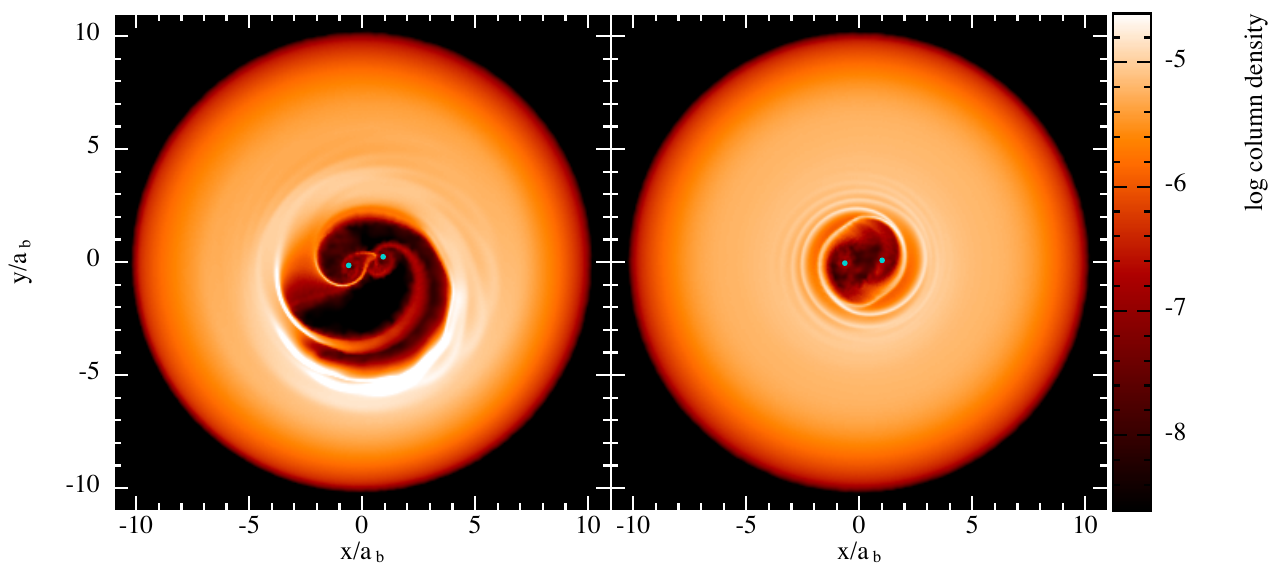}
    \caption{Column density profiles for prograde (left) and retrograde  (right) circumbinary disks  once they have each reached a quasi-steady state. In each panel, the left cyan point represents the primary and the right cyan point represents the secondary. In the prograde case we see a large inner cavity has been evacuated which is punctuated by a stream that feeds the circumprimary disk. The strongest density enhancement in this case occurs at a radius of around $4-5a_{\rm b}$. In the retrograde case the disk extends much closer to the binary with a ring-like structure occurring at around $2a_{\rm b}$ with a thickness of around $0.5a_{\rm b}$. The retrograde disk structure is in better agreement with the observations for DX Cha.}
    \label{fig:both}
\end{figure*}

Once the prograde disk reached the final state, the number of particles in the disk has increased to $\approx 6155000$. In the left panel of  Fig.~\ref{fig:both}, we plot the column density of the steady-state prograde disk when the two stars are at nearly apoapsis (from now on, we redefine time as $t = t' - t_{\rm std}$, where $t'$ denotes the code time and $t_{\rm std}$ is the time when the system reaches a steady state). The axes are given in units of the binary semi-major axis, and the units for the column density color bar are arbitrary. The tidal interaction between the binary and the circumbinary disk results in an eccentric inner cavity with an apocenter of around $\approx 4 a_{\rm b}$ (0.88 au) and the shortest radius $\approx 2 a_{\rm b}$ (0.44 au) to the mass center of the binary. This size agrees with \cite{Hirsh2020} who found a similar cavity size with similar parameters (see their Fig.~9). The most significant density enhancement occurs at a radius of around $\approx 4.5\ a_{\rm b}$ (towards the bottom left in the left panel of Fig.~\ref{fig:both}) which coincides with the apocenter of the eccentric inner edge of the circumbinary disk. There is also a spiral from the inner edge of the circumbinary disk feeding mass into the circumprimary disk.

\begin{figure*}
    \centering
    \includegraphics[width=\linewidth]{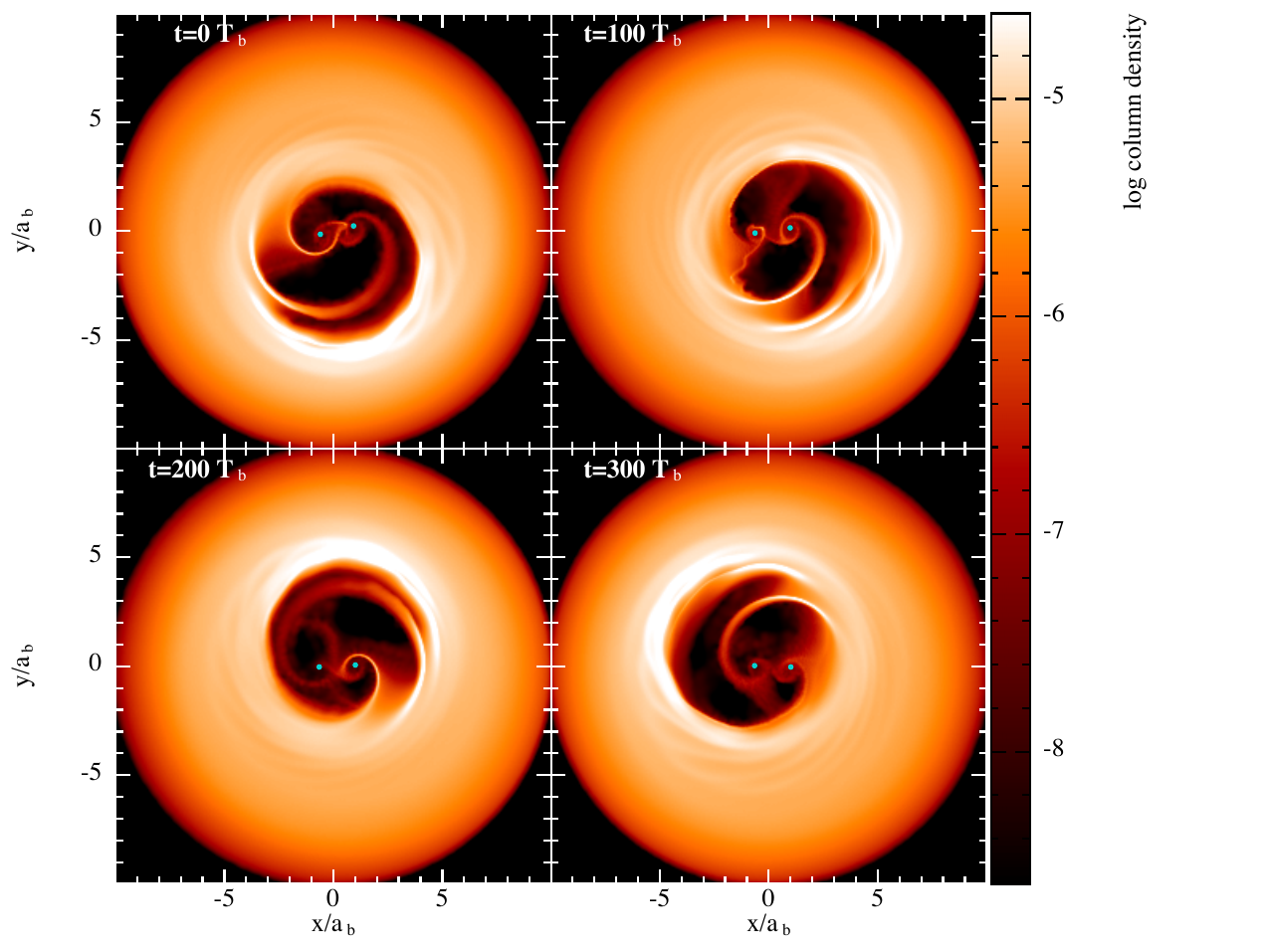}

    \caption{Series of snapshots showing the precession of the density enhancement at the apocentre of the eccentric inner disk edge. The snapshots are spaced by around $100T_{\rm b}$, and the timescale for a full precession of the enhancement is $\approx 500\, T_{\rm b}$ for these parameters. The inner disk structure is not fixed, but varies through the binary orbit; this is shown in more detail in Fig.~\ref{fig:prograde-series}.}
    \label{fig:prograde}
\end{figure*}

The location of the density enhancement at the apocenter of the eccentric inner edge of the circumbinary disk precesses around the binary with time \citep{MacFadyen:2008}. In Fig.~\ref{fig:prograde}, we plot a series of snapshots with time intervals of about 100 $T_{\rm b}$. The position angle of the enhancement region evolves over time and it takes about 500 $T_{\rm b}$ to complete a full revolution around the binary. Throughout this precession, the inner cavity maintains its asymmetric structure and remains at a large distance from the binary.   

\begin{figure*}
    \centering
    \includegraphics[width=\linewidth]{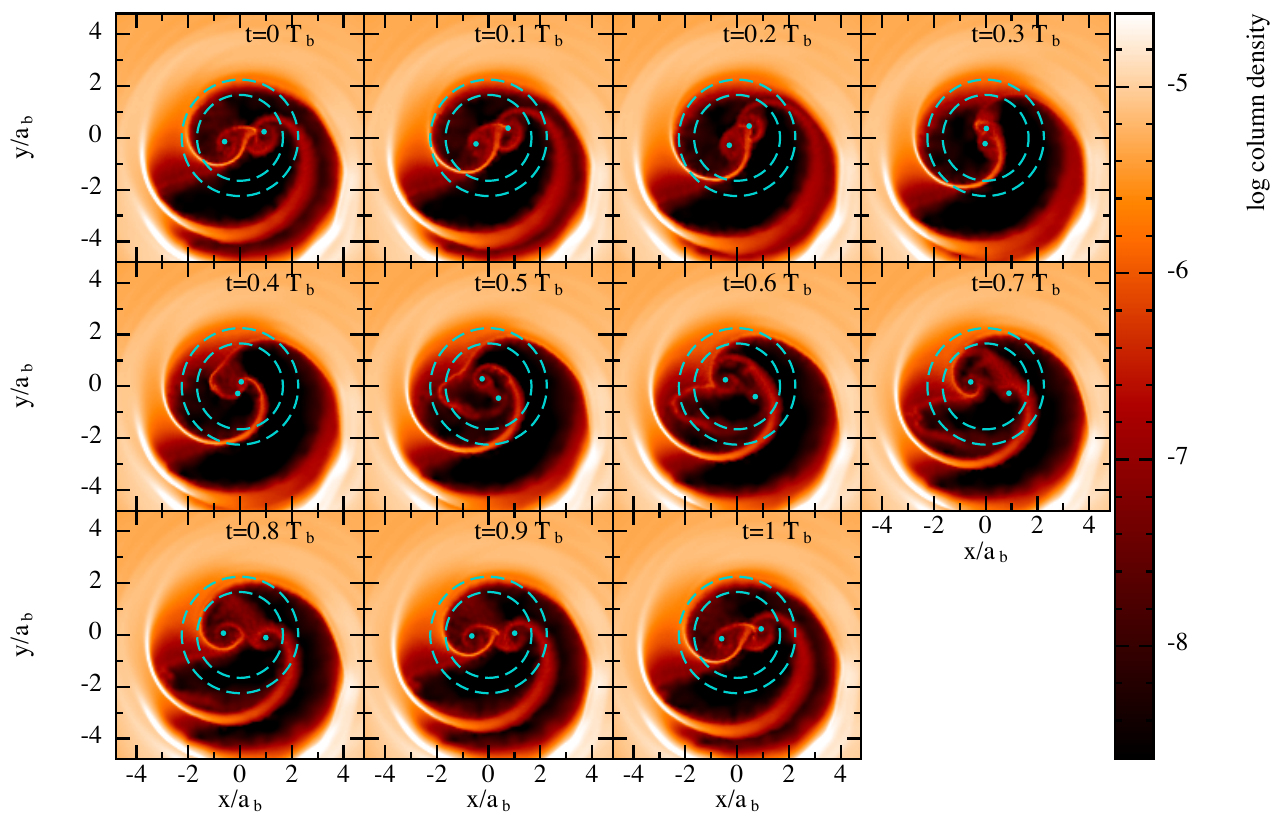}
    \caption{Series of snapshots showing the disk structural changes on short timescales as the binary completes an orbit. Each panel is separated in time by an interval of $0.1 T_{\rm b}$. The two cyan dashed circles indicate the location and FWHM of the azimuthal asymmetric ring modeled in \cite{Juhasz2025}. While the spiral streams can cross the location of the observed ring, the strong density enhancement associated with the inner edge of the circumbinary disk is at a significantly larger radius than the observed ring.}
    \label{fig:prograde-series}
\end{figure*}

To understand the cavity evolution on shorter timescales, we plot in Fig.~\ref{fig:prograde-series} a series of snapshots of the disk structure with an interval of $0.1 T_{\rm b}$. We also mark on the plot with two cyan dashed circles, the location and width (utilizing the full width at half maximum, FWHM) of the ring modeled in \citealt{Juhasz2025}. During the disk evolution, one or two spirals transit through this region, but with significantly lower density than the disk inner edge at much larger radii. 

\subsection{Retrograde case}
Once the retrograde disk reached the final state, the number of particles in the disk has increased to $\approx 5920000$, which is slightly fewer than the prograde case. In the right panel of Fig.~\ref{fig:both}, we plot the column density of the retrograde disk once it has reached a quasi-steady-state (here we also define time $t = 0$ from this snapshot). Compared to the prograde circumbinary disk in the left panel, the retrograde disk has a smaller inner cavity with a radius $\approx 2a_{\rm b}$ and the cavity is also more symmetric, dominated by an $m=2$ mode (as opposed to the prograde case which is dominated by an $m=1$ mode; see \citealt{Nixon2015} for details). 

The smaller cavity is as expected for the retrograde case in which the tidal torque from the binary is significantly weaker \citep[see the figures in][]{Nixonetal2011a,Nixon2012,Nixon2015}. Our results are similar to the lower-left panel of Figure~6 in \citet{Nixon2015} as this is the closest match to the parameters employed here. Looking at Figures 4-6 in \citet{Nixon2015}, we can see that for small $q \leq 0.3$ there is a single prominent arc, while for larger $q$ the structure is comprised of two arcs. 

\begin{figure*}
    \centering
    \includegraphics[width=\linewidth]{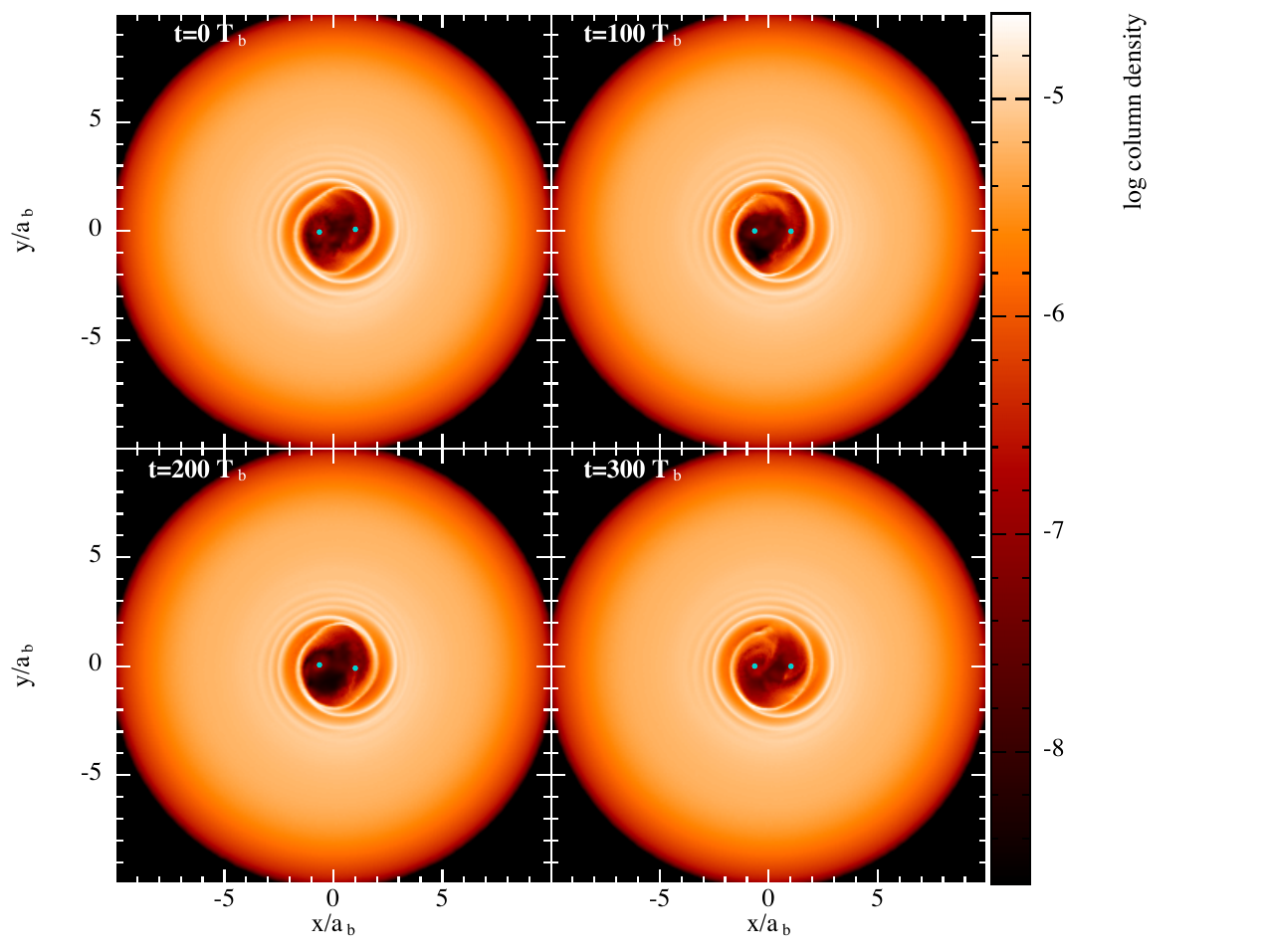}
    \caption{Same as Fig.~\ref{fig:prograde} except for the retrograde, rather than prograde, case.}
    \label{fig:retro}
\end{figure*}

In Fig.~\ref{fig:retro}, which is the same as Fig.~\ref{fig:prograde} but for the retrograde case, we show that the disk structure in the retrograde case does not change substantially over long timescales. In this case the arcs are formed from material that is accelerated by interaction with the stars, predominantly at apocenter; this means that the phase angle of the arcs does not change significantly with time. 

\begin{figure*}
    \centering
    \includegraphics[width=\linewidth]{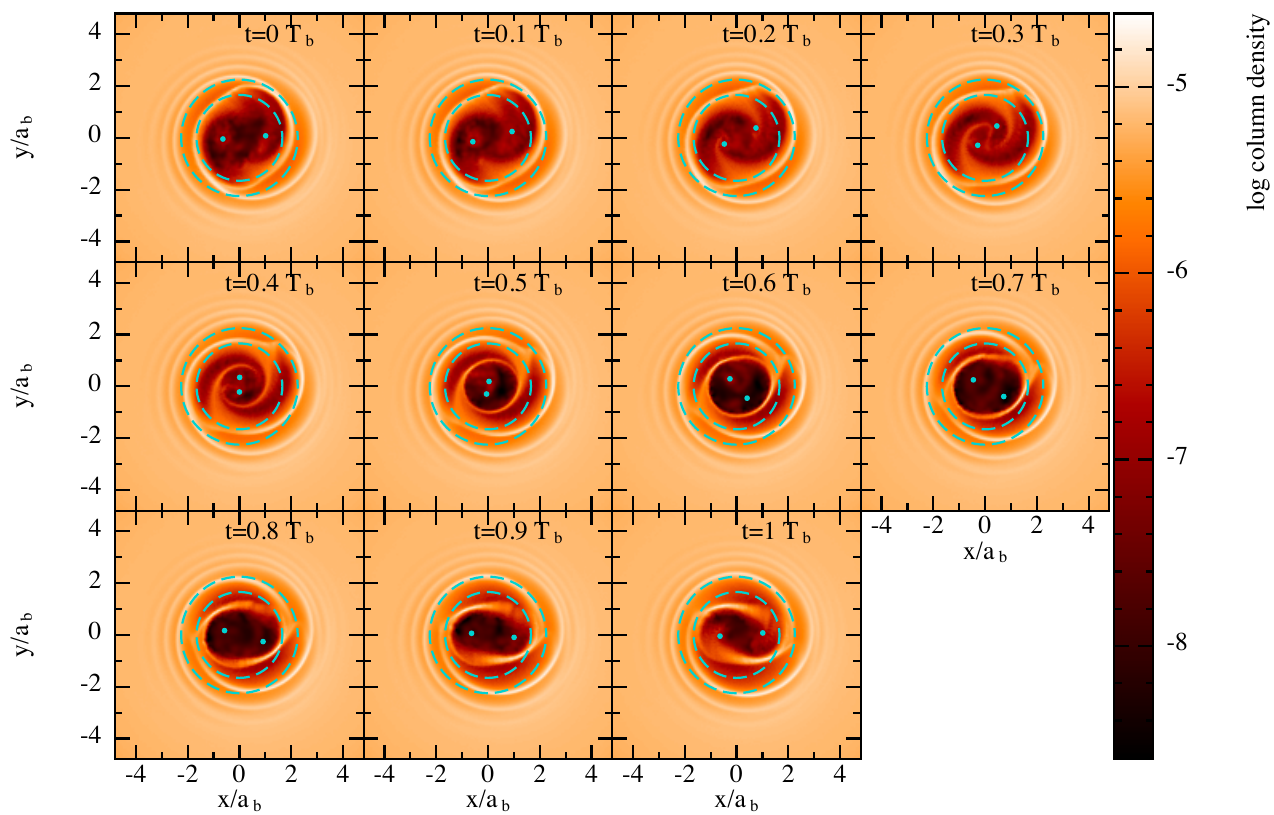}
    \caption{Same as Fig.~\ref{fig:prograde-series} except for the retrograde, rather than prograde, case.}
    \label{fig:retro-series}
\end{figure*}

As in Fig.~\ref{fig:prograde-series} for the prograde case, we also plot in Fig.~\ref{fig:retro-series} the retrograde case throughout a binary orbit with intervals between snapshots of $0.1T_{\rm b}$. Again the two cyan dashed circles represent the location and FWHM of the ring modeled by \cite{Juhasz2025}. Two approximately symmetric spirals with a phase difference of $\approx 180^{\circ}$ are launched outward from the material interacting directly with the binary. The innermost material encroaches right on to the binary, while the density peak in the arcs occurs at around $2a_{\rm b}$ (0.44\,au). This simulation is consistent with the analytic model of \cite{Nixon2015}, which predicts that the strongest dimensionless mode $S_{-1,2}$ occurs at $\approx 2 a_{\rm b}$. The spiral arcs observed in the retrograde circumbinary disk simulation occupy the general area indicated by the data as modeled by \cite{Juhasz2025}.

\section{Discussion}\label{dis}

\subsection{Prograde or retrograde?}
The size and structure of the cavity around the binary is strongly affected by whether the circumbinary disk is prograde or retrograde. Previous simulations of prograde circumbinary disks conclude that the cavity size is $\sim 3-5 a_{\rm b}$ for values of $q_{\rm b}$ and $e_{\rm b}$ that are similar to those of DX Cha \citep[e.g.][]{Thun2016, Hirsh2020, Penzlin2024}. This translates to a size of $\sim 1$\,au, which is significantly larger than implied by the observations \citep[e.g.\@][]{Juhasz2025}.

Indeed, \citet{Dunhill2015} explicitly modeled DX Cha and found a cavity size that is too large for this system. We have also simulated a prograde circumbinary disk with larger $\alpha_{\rm SS}$ = 0.1 to see if a larger disk viscosity can prevent the formation of a large inner cavity. However, the cavity in our simulation is still strongly asymmetric and significantly wider than observed. Moreover, two asymmetric spirals cannot explain the azimuthally asymmetric narrow ring structure modeled from  observations. The density enhancement region located at 4.5 $a_{\rm b}$ ($\simeq$ 1 au) is still too far from the ring's location. Additionally, the period of the position angle of the enhancement region is more than 400 $T_{\rm b}$. This precession rate of the disk $\dot{\omega}$ at $R$ = 4.5 $a_{\rm b}$ is similar to the calculation in \cite{Leung2013} for a test particle orbiting around a binary that 

\begin{equation}
    \dot{\omega} \simeq \frac{3}{4}\frac{M_1M_2}{(M_1+M_2)^2} \left( \frac{a_{\rm b}}{R} \right)^2 \left(1 + \frac{e_{\rm b}}{2} \right)^2.
\end{equation}

With the parameters from our simulation, we found $\dot{\omega} \simeq$ 0.75$^{\circ}$ per $T_{\rm b}$ and thus it takes about 480 $T_{\rm b}$ for the enhancement region to complete one orbit. On the other hand, in Figure 3 of \cite{Juhasz2025}, they found that the ring varies its position angle very fast between their observational epoch 1 and 2 \citep[see equations 4 and 5 of][]{Juhasz2025}. Consequently, neither the spirals nor the density enhancement region can fit the ring structure for the prograde disk.


Alternatively, the disk in DX Cha might be prograde but possess parameters significantly distinct from those we have employed here. Specifically, \cite{Heath2020} demonstrated that prograde disks with large $H/R$ = 0.2 and high viscosity $\alpha_{\rm ss}$ = 0.3 can avoid resonance-driven truncation, allowing the disk to extend much closer to the binary. However, in this scenario, we should observe two prominent symmetric spirals and an azimuthally symmetric rings at the inner edge of circumbinary disk. Moreover, achieving such a large $H/R$ requires extreme physical conditions. Since $H/R \approx c_{\rm s}/v_{\rm k}$, a high $H/R$ implies a very high sound speed relative to the Keplerian velocity, $v_{\rm k}$. For the DX Cha system ($M_{\rm b} = 3.6$\ M$_{\odot}$), at a distance of $R\approx 1$ au, $v_{\rm k}$ is $\sim$ 56 km/s. To sustain $H/R \approx 0.2$, $c_{\rm s}$ must be 11 km/s. Even if we consider that the inner disk region is hot and fully ionized, resulting in a low mean molecular weight of $\mu \approx 0.6$ (as noted by \citealt{Martin2019}) and we assume $m_{\rm p}$ to be the proton mass and the Boltzmann constant is $k_{\rm b}$, the required temperature would be T $\approx\mu m_{\rm p} c_{\rm s}^2/k_{\rm b}\approx 9000$ K. This temperature is significantly higher than typical protoplanetary disk temperatures at 1 au, which are generally limited by dust sublimation($\sim$1500 K) or stellar irradiation equilibrium. 

Regarding viscosity, we used $\alpha_{\rm ss}$ = 0.1, which is already at the upper end of estimates for fully ionized disks \citep{Martin2019} and motivated partly by numerical convenience. Protoplanetary disks are typically expected to have lower viscosity ( e.g. \citealt{Hartmann1998, Rosotti:2023}), which would result in even larger cavities \citep{Dunhill2015}. Therefore, while a prograde disk with extreme scale height and viscosity could theoretically fit the spatial scale of the cavity, the physical conditions required (T $\sim\ 10^4$ K at 1 au) are unlikely to be sustained in the DX Cha system.

Furthermore, the proximity of the disk inner edge to the binary ($\sim$ 1 au) implies high gas temperatures. Based on our simulation parameters, the temperature in this region is T$\sim$3000 K. At such temperatures, dust sublimation occurs and thermal ionization of alkali metals becomes important, likely activating the Magnetorotational Instability (MRI; e.g. \citealt{Jankovic2021,Jankovic2022}). While we do not explicitly simulate magnetohydrodynamics (MHD), our choice of a high viscosity $\alpha_{\rm ss}$ effectively captures the angular momentum transport driven by MRI turbulence in this ionized regime \citep{Martin2019}. Previous comparisons between MHD and hydrodynamical simulations of circumbinary disks indicate that the cavity size is largely governed by the balance between binary torques and viscous spreading, yielding similar results when the effective stress is matched \citep[e.g.][]{Shi2012}. Since higher viscosity leads to a smaller cavity \citep{Dunhill2015}, and our high $\alpha_{\rm ss}$ prograde model still produces a cavity significantly larger than observed, explicitly including magnetic field effects is unlikely to reconcile the prograde scenario with observations. Thus, the hydrodynamic approximation with high $\alpha_{\rm ss}$ serves as a robust upper limit for the ability of a prograde disk to encroach upon the binary.

Finally, we address the impact of thermodynamics. Our simulations employ a locally isothermal equation of state (modified from \citealt{Farris2014}), which is a common approximation but neglects radiative cooling and detailed energy transport. Recent studies have explored circumbinary disk evolution with more sophisticated thermodynamics (e.g. \citealt{Sudarshan2022, Pierens2023, Penzlin2024}). These works generally find that while radiative effects can alter the cavity eccentricity and density contrast, they do not drastically reduce the cavity size for prograde binaries with significant eccentricity.

In fact, if radiative cooling leads to a cooler disk (lower $H/R$), the cavity size in the prograde case often increases or remains large because orbital resonances become stronger in thinner disks \citep{Pierens2023}. Conversely, confirming the findings of \cite{Sudarshan2022}, no thermodynamic treatment has yet demonstrated the ability to shrink the inner cavity of a prograde disk around an eccentric binary ($e_{\rm b} \sim 0.665$) to the size inferred for DX Cha ($\approx$ 2 $a_{\rm b}$). Therefore, while including more complex thermal physics might refine the detailed shape of the arcs in our retrograde models, it is unlikely to enable the prograde scenario to match the compact ring structure observed.

Based on our retrograde circumbinary disk simulation, we infer that DX Cha hosts a retrograde circumbinary disk instead of a prograde circumbinary disk for several reasons. First, eccentric resonances in a retrograde circumbinary disk produce a relatively small and symmetric inner cavity, which has similar features to those reported in previous observations. Second, two symmetric spirals propagate from outward, and their surface densities evolve to the brightest values around 2 $a_{\rm b}$ (0.44 au). This result matches the radius of the asymmetric ring modeled in \cite{Juhasz2025}. Further, \cite{Juhasz2025} employed a model with only a first order azimuthal modulation \citep[see][for details]{Lazareff2017}. This means that the model is restricted to only one maximum brightness and one brightness minimum in the opposite direction over the azimuthal range. As the size of this ring is significantly smaller than previous prograde simulations had suggested, \cite{Juhasz2025} concluded that the ring corresponds to material orbiting within the cavity rather than the inner edge of the circumbinary disk. However, Figure 14 in \cite{Lazareff2017} shows that $\chi^2$ for the second order modulation in some disk systems can be smaller than those with the first order modulation. Thus, we suggest that the asymmetric double spiral present in the retrograde case may be a good fit to the data for the azimuthal structure of the ring. On the other hand, given the sparse uv-coverage of the VLTI/MATISSE observations, the image presented in \citet{Juhasz2025} should be interpreted as the image of a best-fitting parametric brightness model constrained by the interferometric observables measured at the available uv points, rather than a unique image reconstruction. Consequently, higher-complexity structures (e.g. multi-armed spirals) are not uniquely required by the available data and would be difficult to justify without substantially improved uv-coverage. In this work, we therefore restrict our comparison to robust, low-order morphological properties (ring radius/width and low-order azimuthal asymmetry) that are directly captured by the adopted parametric models.

Additionally, it is worth mentioning that the time interval between the second and third observations in  \cite{Juhasz2025} (16.02 $T_{\rm b}$) is nearly a full orbital period of the binary. Hence, the rings fitted in Figure 3 of \cite{Juhasz2025} for these two observations are very similar. On the other hand, the difference in ring positions between the first and second observations in the same figure is due to the time interval between these two observations not being close to a full $T_{\rm b}$. As a result, they fitted two different rings in these two observations. Moreover, the phase angle difference between two rings in the first and second observations is about 180$^{\circ}$ which is consistent with our simulation result.  

\begin{figure*}
    \centering
    \includegraphics[width=0.7\linewidth]{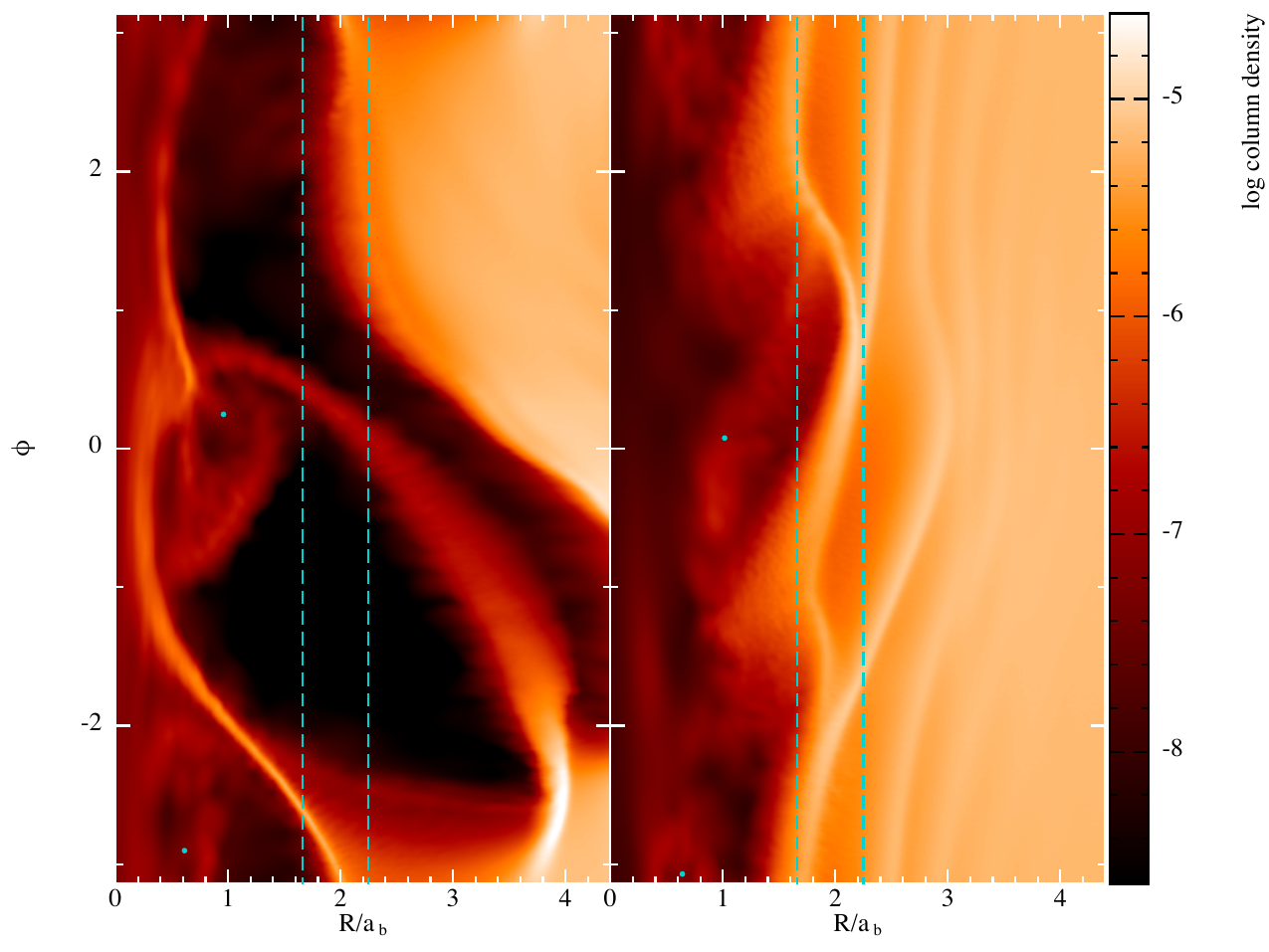}
    \caption{Same as Fig.~\ref{fig:both} except viewed in polar coordinates, $R,\phi$, and showing a smaller range in $R$. The prograde case is on the left and the retrograde case is on the right. The two cyan vertical lines indicate the FWHM of the azimuthal asymmetric ring modeled in \cite{Juhasz2025}.}
    \label{fig:phase}
\end{figure*}

Finally, according to the fitting parameters in \cite{Juhasz2025}, the amplitudes of the azimuthal modulation in three observation epochs are ($A_{mod1}$, $A_{mod2}$, $A_{mod3}$) = (0.77, 0.47, 0.98), respectively. These three values indicate the degree of asymmetry of rings they modeled and the ratio of the brightness minimum to the maximum brightness of rings can be calculated by (1-$A_{mod}$)/(1+$A_{mod}$,). Hence, the ratios for the three observations are 0.13, 0.36, and 0.01, respectively. Because the column density can be used to infer the brightness, in Fig.~\ref{fig:phase}, we convert Fig.~\ref{fig:both} into polar coordinates, and the two cyan vertical lines represent the FWHM of the ring reported in \cite{Juhasz2025}. For the prograde case on the left panel, the ratio of the brightness minimum to the maximum brightness within the region of two lines is $\ll 10^{-2}$ due to the deeper cavity while the ratio of the retrograde case is $\gtrsim 10^{-2}$ principally due to the shallow cavity. Consequently, the retrograde simulation provides a better explanation of the degree of asymmetry measured by \cite{Juhasz2025} than the prograde one.

As the orbital direction of the DX Cha binary has been confirmed \citep{Garcia2013}, the most straightforward way to determine whether DX Cha hosts a retrograde circumbinary disk is through observational data that reveals the kinematic substructures of the disk \citep[e.g.][]{Teague2021}. For moderately inclined disks with optically thick CO observed at high angular resolution, the emitting surfaces of both the front and back side of the disk can be spatially separated in individual ALMA channels \citep[e.g.\@][]{Pinte:2018a,Pinte:2018b,Law:2022,Izquierdo:2023}. In such cases we can thereby distinguish the near- and far-side of the disk using the ALMA data.  However, this cannot yet be achieved for DX Cha with current ALMA observations. In fact it is particularly difficult given the almost face-on inclination with respect to the observer \citep[$i \sim 20^\circ$;][]{Czekala2019}, in combination with a relatively compact protoplanetary disk \citep[$R_{\rm out} \approx 40$\,au;][]{Stapper2024} meaning that ALMA's longest baselines (nominal configuration 10) would be required to angularly resolve this separation. Scattered light observations are equally challenged by the face-on inclination of the source, in which the amount of reflected light received by the observation will be minimized.

\subsection{Accretion rate of DX Cha} \label{acc}
In our simulations, we turn off the back reaction of the binary from the circumbinary disk in order to prevent the orbital evolution of the binary before the circumbinary disk reaches a steady state. To study the orbital evolution of DX Cha and the binary precession due to the disk, we need to turn on the switch to allow the binary to feel the circumbinary disk. However, the disk mass will play a role in affecting these circumbinary disk-induced evolutions. 

Therefore, to set the value of the disk mass in our simulations, $M_{\rm d}$, more precisely, we fit the average accretion rate measured from 93 spectroscopic observations of DX Cha's main component collected from the ESO (European Southern Observatory) archive. We followed the methodology described in \cite{Iglesias2023}, where the mass accretion rate is estimated from the intensity of the H$\alpha$ line emission at 6562.85\AA. The net H$\alpha$ emission intensity can be quantified in terms of its equivalent width as EW$_{\rm emi}$ = EW$_{\rm obs}$ - EW$_{\rm abs}$, where EW$_{\rm obs}$ is the EW of the observed H$\alpha$ spectrum, and EW$_{\rm abs}$ is the EW of the intrinsic H$\alpha$ absorption. We fit a Kurucz model to the wings of the H$\alpha$ line to estimate the EW$_{\rm abs}$. The best-fit model was achieved for $T_{\rm eff}=7250$\,K, $\log (g)=4.6$\,dex, and $v\sin i= 20$\,km s$^{-1}$, adopting solar metallicity and turbulent velocity of 2.0 km s$^{-1}$. The net H$\alpha$ emission flux is then estimated as $F_{\rm emi}$ = EW$_{\rm emi}\times F_{\lambda}$, where $F_{\lambda}$ is the expected flux at the center of the H$\alpha$ line. We estimated $F_{\lambda}$ by using the Kurucz model and scaling it by $(D/R_{\star})^2$, where $D=106.6$\,pc is the distance to the star \citep{GaiaCollab2020} and $R_{\star}=2.67 R_{\odot}$ is the main component's stellar radius, estimated following \cite{Iglesias2023}. Then, the emission line luminosity $L_{\rm emi}$ is calculated as $L_{\rm emi} = 4\pi D^{2}F_{\rm emi}$, and the accretion luminosity $L_{\rm acc}$ is calculated in terms of $L_{\rm emi}$ by following the relationship from \cite{Mendigutia2011b}:
\begin{equation}
    \log \left(\frac{L_{\rm acc}}{L_{\odot}}\right) = A + B\times\log \left(\frac{L_{\rm emi}}{L_{\odot}}\right)
\end{equation}
where $A$ = 2.09 $\pm$ 0.06 and $B$ = 1.00 $\pm$ 0.05 are the relationship values for H$\alpha$ obtained by \cite{Fairlamb2017}. Finally, following \citet{Wichittanakom2020}, the mass accretion rate $\dot M_{\rm acc}$ is determined from $L_{\rm acc}$, $R_{\star}$, and stellar mass $M_{\star}$ as:

\begin{equation}
    \dot M_{\rm acc} = \frac{L_{\rm acc}R_{\star}}{G M_{\star}}
\end{equation}

\begin{figure*}
    \centering
    \includegraphics[width=0.48\linewidth]{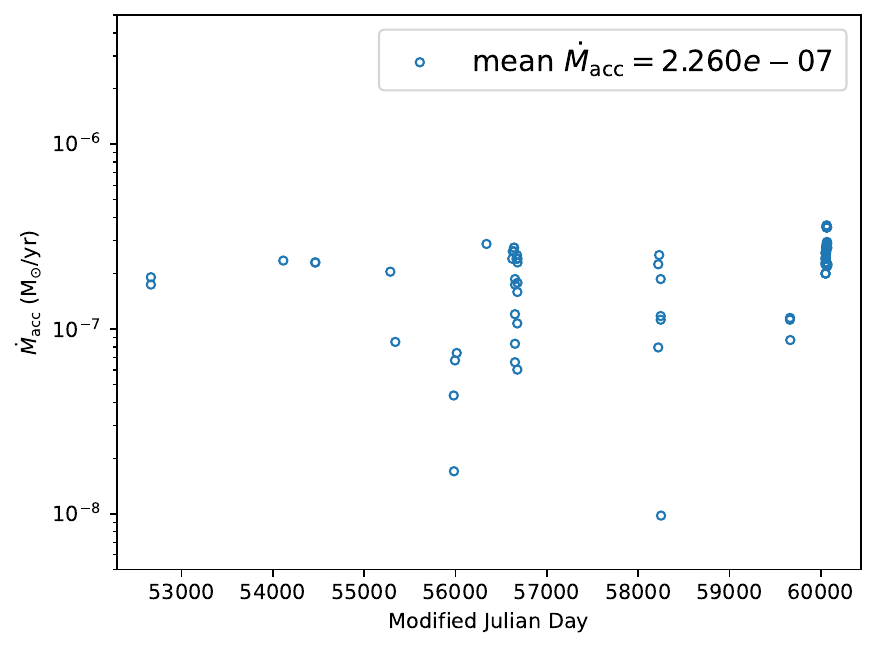}
    \includegraphics[width=0.49\linewidth]{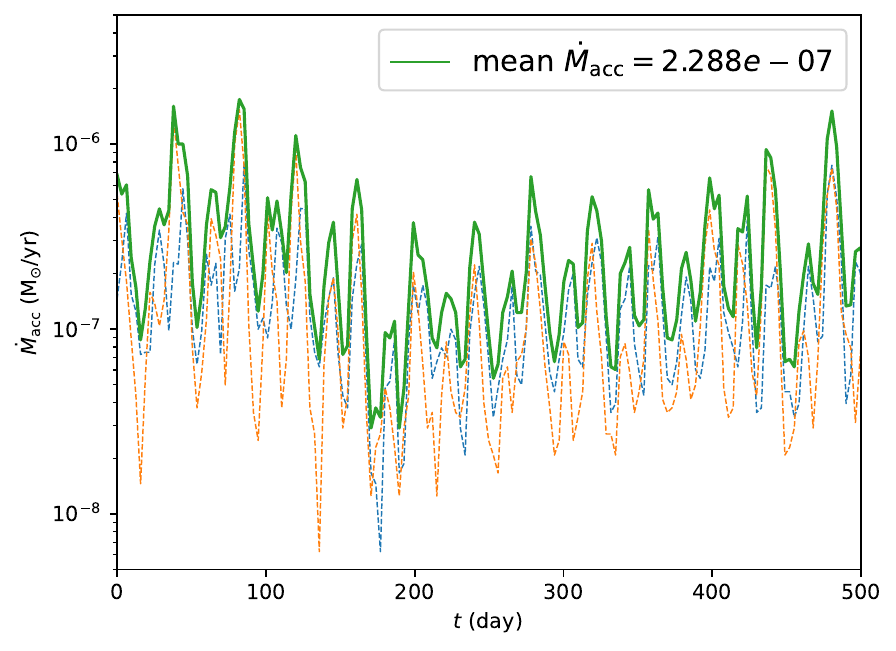}

    \caption{{\it Left}: The fitting results of $\dot M_{\rm acc}$ of DX Cha across the 93 epochs  shown as empty circles. {\it Right}: The new $\dot M_{\rm acc}$ of DX Cha with the time in our simulations after we reduce the disk mass in our model from the original value given in the setup (Section~\ref{num}) by a factor of $\approx 200$. The blue and orange dashed lines represent $\dot M_{\rm acc}$ of the primary and the secondary of DX Cha in the retrograde simulation, respectively. The green line is the total $\dot M_{\rm acc}$.}
    \label{fig:oacc}
\end{figure*}
In the left panel of Fig~\ref{fig:oacc}, we show $\dot M_{\rm acc}$ epochs, which varies widely, ranging from $\sim10^{-8}$ M$_{\odot}$yr$^{-1}$ to $\sim4\times10^{-7}$ M$_{\odot}$yr$^{-1}$. We take the average value of $2.26\times10^{-7}$ M$_{\odot}$yr$^{-1}$. This accretion rate is consistent with typical values observed for Herbig Ae/Be stars \citep{Wichittanakom2020}.

For our retrograde simulations, $\dot M_{\rm acc}$ at a steady state is about $\sim4.76\times10^{-5}$ M$_{\odot}$yr$^{-1}$. Hence, we divide both $M_{\rm d}$ and $\dot{M}_{\rm add}$ by a factor of 200 and restart the simulation for more than 1000 days. In the right panel of Fig~\ref{fig:oacc}, we show that the accretion rate with the time of the primary (blue dashed line) and the secondary (yellow dashed line) of DX Cha with the time after we changed $M_{\rm d}$ in first 500 days. The green line represents the sum of the blue line and the yellow line, and the mean total $\dot M_{\rm acc}$ is $\sim $2.29$\times10^{-7}$\ M$_{\odot}$yr$^{-1}$, which is comparable to the mean value in the left panel.

\subsection{Orbital evolution of DX Cha}
\begin{figure*}
    \centering
    \includegraphics[width=0.48\linewidth]{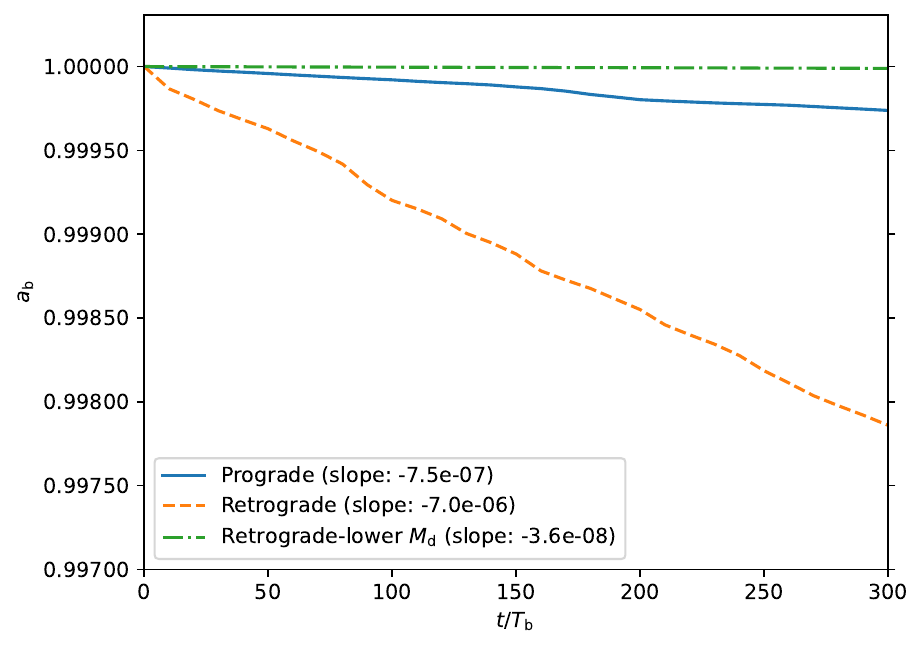}
    \includegraphics[width=0.48\linewidth]{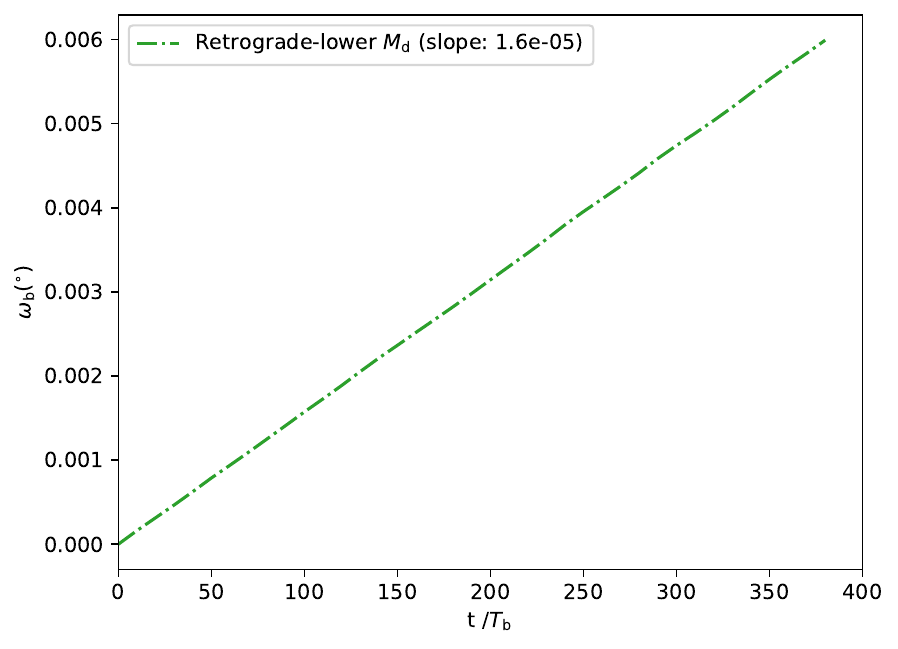}
    \caption{{\it Left}: $a_{\rm b}$ versus $T_{\rm b}$ for prograde (blue), retrograde (orange dashed) and retrograde with the lower $M_{\rm d}$ (green dot-dashed) cases. The slopes in the legend indicate the $\dot{a_{\rm b}}$ rates for three simulations. {\it Right}: Time evolution of $\omega_{\rm b}$ for the retrograde disk simulation.}
    \label{fig:live}
\end{figure*}

In Fig.~\ref{fig:live}, we plot $a_{\rm b}$ versus time for prograde (blue), retrograde (yellow dashed) and retrograde with lower $M_{\rm d}$ (green dot-dashed) cases. All of these three cases show that the binary's orbit shrinks over time. However, the $\dot{a}_{\rm b}$ of retrograde case is faster than the prograde case by an order of magnitude. Because particles in the retrograde circumbinary disk provide the negative specific angular momentum as they accrete onto a binary star, it can speed up the shrinkage rate of the binary \citep{Nixonetal2011a,Nixon2013,Nixon2015}.   

Furthermore, by comparing slopes of yellow dashed and green dot-dashed lines, we confirm that $\dot{a_{\rm b}}$ can be scaled down by $M_{\rm d}$ in our simulations. Hence, if we convert the time unit into years, our results suggest that DX Cha may merge within 2 million years. 

However, in addition to the binary-disk interaction, the tidal interaction between the two stars must be considered because the binary separation is only 0.22 au. The $n$-body simulation in \cite{Chen2024b} shows that an eccentric binary with $e_{\rm b}$ = 0.8 and $a_{\rm b}$ = 0.2 au will merge eventually within 120,000 $T_{\rm b}$. Importantly, in their simulation, they employed the constant time lag approximation, which does not evolve the spins of either star \citep[][\S~3.1]{Baronett2022}. On the other hand, the observation of the young ($\sim 10^6$ yr) Orion OBIc/d association rotation periods of \citep{Stassun1999} show that the majority of rotation periods of those stars are $\leq$ 8 days which is much smaller than the orbital period of DX Cha ($\sim$ 20 days). As a result, the tidal interaction of DX Cha would cause the binary to expand over time if two stars are not tidally locked. Thus, to investigate this topic, we will (in the future) perform $n$-body simulations with the spin evolution for both stars. Combining the binary-disk interaction and the tidal interaction of the binary, we will be able to determine the net effect of the orbital evolution of DX Cha and estimate the shrinkage or expansion rate correctly.

Furthermore, since the phase angle of the spiral in the retrograde circumbinary disk depends on the argument of periapsis of the binary ($\omega_{\rm b}$), it can vary with time due to interactions with the circumbinary disk. Therefore, we plot the time evolution of $\omega_{\rm b}$ in the right panel of Fig.~\ref{fig:live} which shows that the variation of $\omega_{\rm b}$ is $<$ 0.006$^{\circ}$ within 400 $T_{\rm b}$. This implies that the evolution of the spiral's phase angle generated from the retrograde circumbinary disk does not vary significantly within several hundred $T_{\rm b}$. Consequently, model fitting results from future VLTI observations may reveal their position angles depending on the observational epochs.  

\subsection{Retrograde circumbinary planet formation}
Currently, only 16 circumbinary planets have been detected by Kepler and TESS databases \citep{Doyle2011, Welsh2012, Orosz2012a, Orosz2012b, Kostov2014, Welsh2015, Li2016, Kostov2016, Kostov2020, Socia2020} and they are all close to coplanar and prograde with the binary orbital plane. However, recent observations reported that a polar planet around the 2MASS J15104786-2818174 binary system  \citep{Baycroft2025}. This finding implies that planet formation can also occur in misaligned circumbinary disks. While polar circumbinary gas disks such as HD 98800 \citep{Kennedy2019}, V773 Tau B \citep{Kenworthy2022} and a polar circumbinary debris disk around the binary 99 Herculis \citep{Kennedy2012,Smallwood2020} have been found, a retrograde circumbinary disk has not yet been discovered. Retrograde circumbinary disks may be rarer than their prograde counterparts in the universe for a variety of reasons, including their formation mechanism, which may require more extreme environments (e.g., substantial late infall of misaligned material or capturing a star with opposing orbital momentum, see \citealt{Kuffmeier2021, Pelkonen2025, Huhn2025} for details), and because the lifetime of retrograde disks may be shorter than similar prograde disks as the material is not held out by resonances in the retrograde case \cite[e.g.,][]{Nixonetal2011a,Nixon2015}.

Planet formation in a retrograde circumbinary disk should be similar to that in a prograde circumbinary disk. In both prograde and retrograde circumbinary disks, a planet located beyond $5a_{\rm b}$ can be quite stable \citep{Doolin2011, Chen20201}. However, due to planet migration (either type I or type II) a circumbinary planet (CBP) could migrate to the inner edge of the circumbinary disk. As a result, a prograde circumbinary planet may be unstable if its semi-major axis $<$ 3 $a_{\rm b}$, although some theories suggest that a CBP within the inner unstable region will move outward to the innermost stable region due to a balance of orbital excitation from the central stars and the gas-drag force \citep{Yamanaka2019}.

Compared to planet formation in prograde circumbinary disks, retrograde circumbinary disks can provide a relatively stable environment for planetesimal growth near the inner edge. Furthermore, $n$-body simulations show that massive retrograde CBPs can be more stable than prograde CBPs in low and moderate $e_{\rm b}$ systems \cite[e.g.\@][]{Chen20201}. Hence, if we detect a CBP between 1.5 to 2.5 $a_{\rm b}$ in a binary system with a low or middle $e_{\rm b}$, on stability grounds we can expect that it is on a retrograde orbit. On the other hand, although our simulations suggest that a tenuous mini-disk could form around one of stars, its lifetime is quite short. Furthermore, there is no observational evidence of mini-disks in the DX Cha system potentially due to the effects of both stellar wind and stellar magnetic field \citep{Garcia2013}. Hence, we infer that our simulations are not able to answer whether the DX cha system host a mini disk and more complicated models are required to deal with this problem. 

\section{Conclusion}
\label{con}
In this study, we perform SPH simulations to investigate whether the observed structures of the DX Cha circumbinary disk are more consistent with prograde or retrograde rotation of the disk. When the disk is prograde, we find an asymmetric inner cavity extending out to several times the binary semi-major axis, with only low density streams making it to the binary orbit. The strongest density enhancement occurs in a region located at $\approx 4.5 a_{\rm b}$, and which has a precession time of about $500 T_{\rm b}$. This is too large a radial scale to be able to explain the asymmetric ring structure observed for DX Cha by VLTI/MATISSE observations (see left panel of Fig.~\ref{fig:both}).

On the other hand, for the retrograde case, the disk extends much closer to the binary orbit and forms a more symmetric inner cavity that might appear as an asymmetric ring in current analysis of the available observational data. The disk inner edge is located almost exactly where the observations indicate; two spiral arcs are launched and the density peaks at around $2 a_{\rm b}$ (0.44 au; see right panel of Fig.~\ref{fig:both}). Moreover, in both cases the precession of the high-density features takes place on significantly longer timescales than inferred from changes in the phase of the asymmetric ring reported \cite{Juhasz2025}; while this cannot be explained with a prograde disk or through precession of the binary orbit in the retrograde case, it might be explained by the sub-orbital timescale evolution of the arcs formed around the binary in the retrograde case (see Fig.~\ref{fig:retro-series}).

We therefore suggest that DX Cha hosts a retrograde circumbinary disk. This claim can be verified, or falsified, by future observations that confirm the direction of the disk's rotation vector. This can be done, for example, using spatially and velocity resolved kinematics from molecular lines \citep[e.g.][]{Calcino2024}. However, achieving sufficient spatial resolution for this system may be challenging.

We therefore draw the conclusion that detailed observations of compact circumbinary disks around young stellar binaries are an excellent way of advancing our understanding of the physics of disk-binary interaction, which plays a fundamental role in many astrophysical systems from star and planet formation to evolution in stellar mass compact object binaries to supermassive black hole mergers.

Additionally, the environment of DX Cha may play a role in determining the formation of any planets in the circumbinary disk. \cite{Grady2004} show that there are several nearby stars indicating that DX Cha may have undergone a flyby that could truncate or disrupt the circumbinary disk affecting the disk's ability to form planets or the orbits of any planets that have already formed.

Finally, to investigate the orbital interaction between the binary and the disk, we rescaled the mass of the disk by fitting the H$\alpha$ flux of DX Cha across 93 observational epochs. With the rescaled disk mass, the binary-disk interaction leads to the binary's orbit shrinking over time, and the binary may merge within 2 million years \cite[cf.][]{Chen2024b}. However, DX Cha consists of pre-main sequence stars, which usually have faster spin rates than their orbital period. Hence, to determine the real orbital evolution of the binary, the tidal interactions of the two stars have to be taken into account.


\begin{acknowledgements}
We thank Isaac Radley, John Ilee, Catherine Walsh and Melvin Hoare for useful discussions. CC, CJN and DI acknowledge support from the Science and Technology Facilities Council (grant numbers ST/Y000544/1 and ST/X001016/1). CC acknowledges support from the CITA National Fellowship (grant number DIS-2022-568580). CJN acknowledges support from the Leverhulme Trust (grant number RPG-2021-380). This project has received funding from the European Union’s Horizon 2020 research and innovation program under the Marie Skłodowska-Curie grant agreement No 823823 (Dustbusters RISE project). We acknowledge the use of resources provided by the Isambard 3 Tier-2 HPC Facility. Isambard 3 is hosted by the University of Bristol and operated by the GW4 Alliance (https://gw4.ac.uk) and is funded by UK Research and Innovation; and the Engineering and Physical Sciences Research Council [EP/X039137/1].
\end{acknowledgements}

\bibliography{main}

@ARTICLE{Izquierdo:2023,
       author = {{Izquierdo}, A.~F. and {Testi}, L. and {Facchini}, S. and {Rosotti}, G.~P. and {van Dishoeck}, E.~F. and {W{\"o}lfer}, L. and {Paneque-Carre{\~n}o}, T.},
        title = "{The Disc Miner. II. Revealing gas substructures and kinematic signatures from planet-disc interaction through line profile analysis}",
      journal = {\aap},
         year = 2023,
        month = jun,
       volume = {674},
          eid = {A113},
        pages = {A113},
          doi = {10.1051/0004-6361/202245425},
archivePrefix = {arXiv},
       eprint = {2304.03607},
 primaryClass = {astro-ph.EP},
       adsurl = {https://ui.adsabs.harvard.edu/abs/2023A&A...674A.113I}
}

@ARTICLE{Law:2022,
       author = {{Law}, Charles J. and {Crystian}, Sage and {Teague}, Richard and {{\"O}berg}, Karin I. and {Rich}, Evan A. and {Andrews}, Sean M. and {Bae}, Jaehan and {Flaherty}, Kevin and {Guzm{\'a}n}, Viviana V. and {Huang}, Jane and {Ilee}, John D. and {Kastner}, Joel H. and {Loomis}, Ryan A. and {Long}, Feng and {P{\'e}rez}, Laura M. and {P{\'e}rez}, Sebasti{\'a}n and {Qi}, Chunhua and {Rosotti}, Giovanni P. and {Ru{\'\i}z-Rodr{\'\i}guez}, Dary and {Tsukagoshi}, Takashi and {Wilner}, David J.},
        title = "{CO Line Emission Surfaces and Vertical Structure in Midinclination Protoplanetary Disks}",
      journal = {\apj},
         year = 2022,
        month = jun,
       volume = {932},
       number = {2},
          eid = {114},
        pages = {114},
          doi = {10.3847/1538-4357/ac6c02},
archivePrefix = {arXiv},
       eprint = {2205.01776},
 primaryClass = {astro-ph.EP},
       adsurl = {https://ui.adsabs.harvard.edu/abs/2022ApJ...932..114L}
}

@ARTICLE{Pinte:2018b,
       author = {{Pinte}, C. and {Price}, D.~J. and {M{\'e}nard}, F. and {Duch{\^e}ne}, G. and {Dent}, W.~R.~F. and {Hill}, T. and {de Gregorio-Monsalvo}, I. and {Hales}, A. and {Mentiplay}, D.},
        title = "{Kinematic Evidence for an Embedded Protoplanet in a Circumstellar Disk}",
      journal = {\apjl},
         year = 2018,
        month = jun,
       volume = {860},
       number = {1},
          eid = {L13},
        pages = {L13},
          doi = {10.3847/2041-8213/aac6dc},
archivePrefix = {arXiv},
       eprint = {1805.10293},
 primaryClass = {astro-ph.SR},
       adsurl = {https://ui.adsabs.harvard.edu/abs/2018ApJ...860L..13P}
}

@ARTICLE{Pinte:2018a,
       author = {{Pinte}, C. and {M{\'e}nard}, F. and {Duch{\^e}ne}, G. and {Hill}, T. and {Dent}, W.~R.~F. and {Woitke}, P. and {Maret}, S. and {van der Plas}, G. and {Hales}, A. and {Kamp}, I. and {Thi}, W.~F. and {de Gregorio-Monsalvo}, I. and {Rab}, C. and {Quanz}, S.~P. and {Avenhaus}, H. and {Carmona}, A. and {Casassus}, S.},
        title = "{Direct mapping of the temperature and velocity gradients in discs. Imaging the vertical CO snow line around IM Lupi}",
      journal = {\aap},
         year = 2018,
        month = jan,
       volume = {609},
          eid = {A47},
        pages = {A47},
          doi = {10.1051/0004-6361/201731377},
archivePrefix = {arXiv},
       eprint = {1710.06450},
 primaryClass = {astro-ph.SR},
       adsurl = {https://ui.adsabs.harvard.edu/abs/2018A&A...609A..47P}
}

@ARTICLE{MacFadyen:2008,
       author = {{MacFadyen}, Andrew I. and {Milosavljevi{\'c}}, Milo{\v{s}}},
        title = "{An Eccentric Circumbinary Accretion Disk and the Detection of Binary Massive Black Holes}",
      journal = {\apj},
         year = 2008,
        month = jan,
       volume = {672},
       number = {1},
        pages = {83-93},
          doi = {10.1086/523869},
archivePrefix = {arXiv},
       eprint = {astro-ph/0607467},
 primaryClass = {astro-ph},
       adsurl = {https://ui.adsabs.harvard.edu/abs/2008ApJ...672...83M}
}

@ARTICLE{Ivanov:2015,
       author = {{Ivanov}, P.~B. and {Papaloizou}, J.~C.~B. and {Paardekooper}, S. -J. and {Polnarev}, A.~G.},
        title = "{The evolution of a binary in a retrograde circular orbit embedded in an accretion disk}",
      journal = {\aap},
         year = 2015,
        month = apr,
       volume = {576},
          eid = {A29},
        pages = {A29},
          doi = {10.1051/0004-6361/201424359},
archivePrefix = {arXiv},
       eprint = {1410.3250},
 primaryClass = {astro-ph.HE},
       adsurl = {https://ui.adsabs.harvard.edu/abs/2015A&A...576A..29I}
}

@ARTICLE{Kraus:2020,
       author = {{Kraus}, Stefan and {Kreplin}, Alexander and {Young}, Alison K. and {Bate}, Matthew R. and {Monnier}, John D. and {Harries}, Tim J. and {Avenhaus}, Henning and {Kluska}, Jacques and {Laws}, Anna S.~E. and {Rich}, Evan A. and {Willson}, Matthew and {Aarnio}, Alicia N. and {Adams}, Fred C. and {Andrews}, Sean M. and {Anugu}, Narsireddy and {Bae}, Jaehan and {ten Brummelaar}, Theo and {Calvet}, Nuria and {Cur{\'e}}, Michel and {Davies}, Claire L. and {Ennis}, Jacob and {Espaillat}, Catherine and {Gardner}, Tyler and {Hartmann}, Lee and {Hinkley}, Sasha and {Labdon}, Aaron and {Lanthermann}, Cyprien and {LeBouquin}, Jean-Baptiste and {Schaefer}, Gail H. and {Setterholm}, Benjamin R. and {Wilner}, David and {Zhu}, Zhaohuan},
        title = "{A triple-star system with a misaligned and warped circumstellar disk shaped by disk tearing}",
      journal = {Science},
         year = 2020,
        month = sep,
       volume = {369},
       number = {6508},
        pages = {1233-1238},
          doi = {10.1126/science.aba4633},
archivePrefix = {arXiv},
       eprint = {2004.01204},
 primaryClass = {astro-ph.SR},
       adsurl = {https://ui.adsabs.harvard.edu/abs/2020Sci...369.1233K}
}

@ARTICLE{Dunhill:2014,
       author = {{Dunhill}, A.~C. and {Alexander}, R.~D. and {Nixon}, C.~J. and {King}, A.~R.},
        title = "{Misaligned accretion on to supermassive black hole binaries}",
      journal = {\mnras},
         year = 2014,
        month = dec,
       volume = {445},
       number = {3},
        pages = {2285-2296},
          doi = {10.1093/mnras/stu1914},
archivePrefix = {arXiv},
       eprint = {1409.3842},
 primaryClass = {astro-ph.HE},
       adsurl = {https://ui.adsabs.harvard.edu/abs/2014MNRAS.445.2285D}
}

@ARTICLE{Goldreich:1979,
       author = {{Goldreich}, P. and {Tremaine}, S.},
        title = "{The excitation of density waves at the Lindblad and corotation resonances by an external potential.}",
      journal = {\apj},
         year = 1979,
        month = nov,
       volume = {233},
        pages = {857-871},
          doi = {10.1086/157448},
       adsurl = {https://ui.adsabs.harvard.edu/abs/1979ApJ...233..857G}
}

@ARTICLE{Papaloizou:1977,
       author = {{Papaloizou}, J. and {Pringle}, J.~E.},
        title = "{Tidal torques on accretion discs in close binary systems.}",
      journal = {\mnras},
         year = 1977,
        month = nov,
       volume = {181},
        pages = {441-454},
          doi = {10.1093/mnras/181.3.441},
       adsurl = {https://ui.adsabs.harvard.edu/abs/1977MNRAS.181..441P}
}

@ARTICLE{Pringle:1972,
       author = {{Pringle}, J.~E. and {Rees}, M.~J.},
        title = "{Accretion Disc Models for Compact X-Ray Sources}",
      journal = {\aap},
         year = 1972,
        month = oct,
       volume = {21},
        pages = {1},
       adsurl = {https://ui.adsabs.harvard.edu/abs/1972A&A....21....1P}
}

@ARTICLE{Lin:1986,
       author = {{Lin}, D.~N.~C. and {Papaloizou}, J.},
        title = "{On the Tidal Interaction between Protoplanets and the Primordial Solar Nebula. II. Self-Consistent Nonlinear Interaction}",
      journal = {\apj},
         year = 1986,
        month = aug,
       volume = {307},
        pages = {395},
          doi = {10.1086/164426},
       adsurl = {https://ui.adsabs.harvard.edu/abs/1986ApJ...307..395L}
}

@INPROCEEDINGS{Bouvier2007,
       author = {{Bouvier}, J. and {Alencar}, S.~H.~P. and {Harries}, T.~J. and {Johns-Krull}, C.~M. and {Romanova}, M.~M.},
        title = "{Magnetospheric Accretion in Classical T Tauri Stars}",
    booktitle = {Protostars and Planets V},
         year = 2007,
       editor = {{Reipurth}, Bo and {Jewitt}, David and {Keil}, Klaus},
        month = jan,
        pages = {479},
          doi = {10.48550/arXiv.astro-ph/0603498},
archivePrefix = {arXiv},
       eprint = {astro-ph/0603498},
 primaryClass = {astro-ph},
       adsurl = {https://ui.adsabs.harvard.edu/abs/2007prpl.conf..479B}
}

@ARTICLE{Nixon2013,
       author = {{Nixon}, Chris and {King}, Andrew and {Price}, Daniel},
        title = "{Tearing up the disc: misaligned accretion on to a binary}",
      journal = {\mnras},
         year = 2013,
        month = sep,
       volume = {434},
       number = {3},
        pages = {1946-1954},
          doi = {10.1093/mnras/stt1136},
archivePrefix = {arXiv},
       eprint = {1307.0010},
 primaryClass = {astro-ph.HE},
       adsurl = {https://ui.adsabs.harvard.edu/abs/2013MNRAS.434.1946N}
}

@ARTICLE{Aly2015,
author = {{Aly}, H. and {Dehnen}, W. and {Nixon}, C. and {King}, A.},
title = "{Misaligned gas discs around eccentric black hole binaries and implications for the final-parsec problem}",
journal = {MNRAS},
archivePrefix = "arXiv",
eprint = {1501.04623},
primaryClass = "astro-ph.HE",
year = 2015,
month = may,
volume = 449,
pages = {65-76},
doi = {10.1093/mnras/stv128},
adsurl = {http://adsabs.harvard.edu/abs/2015MNRAS.449...65A}
}

@ARTICLE{Artymowicz1994,
author = {{Artymowicz}, P. and {Lubow}, S.~H.},
title = "{Dynamics of binary-disk interaction. 1: Resonances and disk gap sizes}",
journal = {ApJ},
year = 1994,
month = feb,
volume = 421,
pages = {651-667},
doi = {10.1086/173679},
adsurl = {http://adsabs.harvard.edu/abs/1994ApJ...421..651A}
}

@ARTICLE{Artymowicz1996,
author = {{Artymowicz}, P. and {Lubow}, S.~H.},
title = "{Mass Flow through Gaps in Circumbinary Disks}",
journal = {ApJl},
year = 1996,
month = aug,
volume = 467,
pages = {L77},
doi = {10.1086/310200},
adsurl = {http://adsabs.harvard.edu/abs/1996ApJ...467L..77A}
}

@ARTICLE{Bailer-Jones2021,
       author = {{Bailer-Jones}, C.~A.~L. and {Rybizki}, J. and {Fouesneau}, M. and {Demleitner}, M. and {Andrae}, R.},
    title = "{Estimating Distances from Parallaxes. V. Geometric and Photogeometric Distances to 1.47 Billion Stars in Gaia Early Data Release 3}",
    journal = {\aj},
         year = 2021,
        month = mar,
       volume = {161},
       number = {3},
          eid = {147},
        pages = {147},
          doi = {10.3847/1538-3881/abd806},
archivePrefix = {arXiv},
       eprint = {2012.05220},
 primaryClass = {astro-ph.SR},
       adsurl = {https://ui.adsabs.harvard.edu/abs/2021AJ....161..147B}
}

@ARTICLE{BH1991,
author = {{Balbus}, S.~A. and {Hawley}, J.~F.},
title = "{A powerful local shear instability in weakly magnetized disks. I - Linear analysis. II - Nonlinear evolution}",
journal = {ApJ},
year = 1991,
month = jul,
volume = 376,
pages = {214-233},
doi = {10.1086/170270},
adsurl = {http://adsabs.harvard.edu/abs/1991ApJ...376..214B}
}

@ARTICLE{Baronett2022,
       author = {{Baronett}, Stanley A. and {Ferich}, Noah and {Tamayo}, Daniel and {Steffen}, Jason H.},
        title = "{Stellar evolution and tidal dissipation in REBOUNDx}",
      journal = {\mnras},
         year = 2022,
        month = mar,
       volume = {510},
       number = {4},
        pages = {6001-6009},
          doi = {10.1093/mnras/stac043},
archivePrefix = {arXiv},
       eprint = {2101.12277},
 primaryClass = {astro-ph.SR},
       adsurl = {https://ui.adsabs.harvard.edu/abs/2022MNRAS.510.6001B}
}

@ARTICLE{Bate2002,
       author = {{Bate}, M.~R. and {Ogilvie}, G.~I. and {Lubow}, S.~H. and {Pringle}, J.~E.},
        title = "{The excitation, propagation and dissipation of waves in accretion discs: the non-linear axisymmetric case}",
      journal = {\mnras},
         year = 2002,
        month = may,
       volume = {332},
       number = {3},
        pages = {575-600},
          doi = {10.1046/j.1365-8711.2002.05289.x},
archivePrefix = {arXiv},
       eprint = {astro-ph/0201149},
 primaryClass = {astro-ph},
       adsurl = {https://ui.adsabs.harvard.edu/abs/2002MNRAS.332..575B}
}

@ARTICLE{Bate2003,
author = {{Bate}, M.~R. and {Bonnell}, I.~A. and {Bromm}, V.},
title = "{The formation of a star cluster: predicting the properties of stars and brown dwarfs}",
journal = {\mnras},
eprint = {astro-ph/0212380},
year = 2003,
month = mar,
volume = 339,
pages = {577-599},
doi = {10.1046/j.1365-8711.2003.06210.x},
adsurl = {http://adsabs.harvard.edu/abs/2003MNRAS.339..577B}
}

@article{Bate2018,
    author = {Bate, Matthew R},
    title = "{On the diversity and statistical properties of protostellar discs}",
    journal = {\mnras},
    volume = {475},
    number = {4},
    pages = {5618-5658},
    year = {2018},
    month = {01},
    issn = {0035-8711},
    doi = {10.1093/mnras/sty169},
    url = {https://doi.org/10.1093/mnras/sty169},
    eprint = {http://oup.prod.sis.lan/mnras/article-pdf/475/4/5618/24243422/sty169.pdf},
}

@ARTICLE{Baycroft2025,
       author = {{Baycroft}, Thomas A. and {Sairam}, Lalitha and {Triaud}, Amaury H.~M. J and {Correia}, Alexandre C.~M.},
        title = "{Evidence for a polar circumbinary exoplanet orbiting a pair of eclipsing brown dwarfs}",
      journal = {arXiv e-prints},
         year = 2025,
        month = apr,
          eid = {arXiv:2504.12209},
        pages = {arXiv:2504.12209},
          doi = {10.48550/arXiv.2504.12209},
archivePrefix = {arXiv},
       eprint = {2504.12209},
 primaryClass = {astro-ph.EP},
       adsurl = {https://ui.adsabs.harvard.edu/abs/2025arXiv250412209B}
}

@ARTICLE{Bohm2004,
       author = {{B{\"o}hm}, T. and {Catala}, C. and {Balona}, L. and {Carter}, B.},
        title = "{Spectroscopic monitoring of the Herbig Ae star HD 104237.  I. Multiperiodic stellar oscillations}",
      journal = {\aap},
         year = 2004,
        month = dec,
       volume = {427},
        pages = {907-922},
          doi = {10.1051/0004-6361:20041227},
       adsurl = {https://ui.adsabs.harvard.edu/abs/2004A&A...427..907B}
}

@ARTICLE{Brinch2016,
author = {{Brinch}, C. and {J{\o}rgensen}, J.~K. and {Hogerheijde}, M.~R. and 
{Nelson}, R.~P. and {Gressel}, O.},
title = "{Misaligned Disks in the Binary Protostar IRS 43}",
journal = {\apjl},
archivePrefix = "arXiv",
eprint = {1610.03626},
primaryClass = "astro-ph.SR",
year = 2016,
month = oct,
volume = 830,
eid = {L16},
pages = {L16},
doi = {10.3847/2041-8205/830/1/L16},
adsurl = {http://adsabs.harvard.edu/abs/2016ApJ...830L..16B}
}

@ARTICLE{Brittain2023,
       author = {{Brittain}, Sean D. and {Kamp}, Inga and {Meeus}, Gwendolyn and {Oudmaijer}, Ren{\'e} D. and {Waters}, L.~B.~F.~M.},
        title = "{Herbig Stars}",
      journal = {\ssr},
         year = 2023,
        month = feb,
       volume = {219},
       number = {1},
          eid = {7},
        pages = {7},
          doi = {10.1007/s11214-023-00949-z},
archivePrefix = {arXiv},
       eprint = {2301.01165},
 primaryClass = {astro-ph.SR},
       adsurl = {https://ui.adsabs.harvard.edu/abs/2023SSRv..219....7B}
}

@article{Calcino2024,
    author = {Calcino, Josh and Norfolk, Brodie J and Price, Daniel J and Hilder, Thomas and Speedie, Jessica and Pinte, Christophe and Garg, Himanshi and Teague, Richard and Hall, Cassandra and Stadler, Jochen},
    title = {Observational signatures of circumbinary discs - II. Kinematic signatures in velocity residuals},
    journal = {Monthly Notices of the Royal Astronomical Society},
    volume = {534},
    number = {3},
    pages = {2904-2917},
    year = {2024},
    month = {10},
    issn = {0035-8711},
    doi = {10.1093/mnras/stae2233},
    url = {https://doi.org/10.1093/mnras/stae2233},
    eprint = {https://academic.oup.com/mnras/article-pdf/534/3/2904/59811370/stae2233.pdf},
}

@article{Chen20201,
    author = {Chen, Cheng and Lubow, Stephen H and Martin, Rebecca G},
    title = "{Polar planets around highly eccentric binaries are the most stable}",
    journal = {\mnras},
    volume = {494},
    number = {4},
    pages = {4645-4655},
    year = {2020},
    month = {04},
    issn = {0035-8711},
    doi = {10.1093/mnras/staa1037},
    url = {https://doi.org/10.1093/mnras/staa1037},
    eprint = {https://academic.oup.com/mnras/article-pdf/494/4/4645/33174329/staa1037.pdf},
}

@article{Chen2024b,
    author = {Chen, Cheng and Baronett, Stanley A and Nixon, C J and Martin, Rebecca G},
    title = {On the origin of polar planets around single stars},
    journal = {Monthly Notices of the Royal Astronomical Society: Letters},
    volume = {533},
    number = {1},
    pages = {L37-L42},
    year = {2024},
    month = {06},
    issn = {1745-3925},
    doi = {10.1093/mnrasl/slae058},
    url = {https://doi.org/10.1093/mnrasl/slae058},
    eprint = {https://academic.oup.com/mnrasl/article-pdf/533/1/L37/58455405/slae058.pdf},
}

@ARTICLE{Chen2025a,
       author = {{Chen}, Cheng and {Nixon}, C.~J.},
        title = "{Minimizing the numerical viscosity in smoothed particle hydrodynamics simulations of discs}",
      journal = {\mnras},
         year = 2025,
        month = jul,
       volume = {540},
       number = {3},
        pages = {2465-2473},
          doi = {10.1093/mnras/staf881},
archivePrefix = {arXiv},
       eprint = {2505.24343},
 primaryClass = {astro-ph.SR},
       adsurl = {https://ui.adsabs.harvard.edu/abs/2025MNRAS.540.2465C}
}

@ARTICLE{Cowley2013,
       author = {{Cowley}, C.~R. and {Castelli}, F. and {Hubrig}, S.},
        title = "{The Herbig Ae SB2 system HD 104237}",
      journal = {\mnras},
         year = 2013,
        month = jun,
       volume = {431},
       number = {4},
        pages = {3485-3493},
          doi = {10.1093/mnras/stt430},
archivePrefix = {arXiv},
       eprint = {1303.1850},
 primaryClass = {astro-ph.SR},
       adsurl = {https://ui.adsabs.harvard.edu/abs/2013MNRAS.431.3485C}
}

@ARTICLE{Cullen2010,
       author = {{Cullen}, Lee and {Dehnen}, Walter},
        title = "{Inviscid smoothed particle hydrodynamics}",
      journal = {\mnras},
         year = 2010,
        month = oct,
       volume = {408},
       number = {2},
        pages = {669-683},
          doi = {10.1111/j.1365-2966.2010.17158.x},
archivePrefix = {arXiv},
       eprint = {1006.1524},
 primaryClass = {astro-ph.IM},
       adsurl = {https://ui.adsabs.harvard.edu/abs/2010MNRAS.408..669C}
}

@article{Czekala2019,
	doi = {10.3847/1538-4357/ab287b},
	url = {https://doi.org/10.3847%2F1538-4357%2Fab287b},
	year = 2019,
	month = {sep},
	publisher = {American Astronomical Society},
	volume = {883},
	number = {1},
	pages = {22},
	author = {Ian Czekala and Eugene Chiang and Sean M. Andrews and Eric L. N. Jensen and Guillermo Torres and David J. Wilner and Keivan G. Stassun and Bruce Macintosh},
	title = {The Degree of Alignment between Circumbinary Disks and Their Binary Hosts},
	journal = {The Astrophysical Journal},
}

@ARTICLE{Doolin2011,
author = {{Doolin}, S. and {Blundell}, K.~M.},
title = "{The dynamics and stability of circumbinary orbits}",
journal = {\mnras},
archivePrefix = "arXiv",
eprint = {1108.4144},
primaryClass = "astro-ph.SR",
year = 2011,
month = dec,
volume = 418,
pages = {2656-2668},
doi = {10.1111/j.1365-2966.2011.19657.x},
adsurl = {http://adsabs.harvard.edu/abs/2011MNRAS.418.2656D}
}

@ARTICLE{Doyle2011,
author = {{Doyle}, L.~R. and {Carter}, J.~A. and {Fabrycky}, D.~C. and 
{Slawson}, R.~W. and {Howell}, S.~B. and {Winn}, J.~N. and {Orosz}, J.~A. and 
{et al.}},
title = "{Kepler-16: A Transiting Circumbinary Planet}",
journal = {Science},
archivePrefix = "arXiv",
eprint = {1109.3432},
primaryClass = "astro-ph.EP",
year = 2011,
month = sep,
volume = 333,
pages = {1602},
doi = {10.1126/science.1210923},
adsurl = {http://adsabs.harvard.edu/abs/2011Sci...333.1602D}
}

@ARTICLE{Drewes2021,
       author = {{Drewes}, N.~C. and {Nixon}, C.~J.},
        title = "{On the Dynamics of Low-viscosity Warped Disks around Black Holes}",
      journal = {\apj},
         year = 2021,
        month = dec,
       volume = {922},
       number = {2},
          eid = {243},
        pages = {243},
          doi = {10.3847/1538-4357/ac2609},
archivePrefix = {arXiv},
       eprint = {2106.11090},
 primaryClass = {astro-ph.HE},
       adsurl = {https://ui.adsabs.harvard.edu/abs/2021ApJ...922..243D}
}

@ARTICLE{Dunhill2015,
author = {{Dunhill}, A.~C. and {Cuadra}, J. and {Dougados}, C.},
title = "{Precession and accretion in circumbinary discs: the case of HD 104237}",
journal = {\mnras},
archivePrefix = "arXiv",
eprint = {1411.0687},
primaryClass = "astro-ph.SR",
year = 2015,
month = apr,
volume = 448,
pages = {3545-3554},
doi = {10.1093/mnras/stv284},
adsurl = {http://adsabs.harvard.edu/abs/2015MNRAS.448.3545D}
}

@ARTICLE{Eggleton1983,
author = {{Eggleton}, P.~P.},
title = "{Approximations to the radii of Roche lobes}",
journal = {ApJ},
year = 1983,
month = may,
volume = 268,
pages = {368},
doi = {10.1086/160960},
adsurl = {http://adsabs.harvard.edu/abs/1983ApJ...268..368E}
}

@ARTICLE{Fairlamb2017,
       author = {{Fairlamb}, J.~R. and {Oudmaijer}, R.~D. and {Mendigutia}, I. and {Ilee}, J.~D. and {van den Ancker}, M.~E.},
        title = "{A spectroscopic survey of Herbig Ae/Be stars with X-Shooter - II. Accretion diagnostic lines}",
      journal = {\mnras},
         year = 2017,
        month = feb,
       volume = {464},
       number = {4},
        pages = {4721-4735},
          doi = {10.1093/mnras/stw2643},
archivePrefix = {arXiv},
       eprint = {1610.09636},
 primaryClass = {astro-ph.SR},
       adsurl = {https://ui.adsabs.harvard.edu/abs/2017MNRAS.464.4721F}
}

@ARTICLE{Farris2014,
       author = {{Farris}, Brian D. and {Duffell}, Paul and {MacFadyen}, Andrew I. and {Haiman}, Zoltan},
        title = "{Binary Black Hole Accretion from a Circumbinary Disk: Gas Dynamics inside the Central Cavity}",
      journal = {\apj},
         year = 2014,
        month = mar,
       volume = {783},
       number = {2},
          eid = {134},
        pages = {134},
          doi = {10.1088/0004-637X/783/2/134},
archivePrefix = {arXiv},
       eprint = {1310.0492},
 primaryClass = {astro-ph.HE},
       adsurl = {https://ui.adsabs.harvard.edu/abs/2014ApJ...783..134F}
}

@ARTICLE{GaiaCollab2020,
       author = {{Gaia Collaboration} and {Brown}, A.~G.~A. and {Vallenari}, A. and {Prusti}, T. and {de Bruijne}, J.~H.~J. and {Babusiaux}, C. and {Biermann}, M. and {Creevey}, O.~L. and {Evans}, D.~W. and {Eyer}, L. and {Hutton}, A. and {Jansen}, F. and {Jordi}, C. and {Klioner}, S.~A. and {Lammers}, U. and {Lindegren}, L. and {Luri}, X. and {Mignard}, F. and {Panem}, C. and {Pourbaix}, D. and {Randich}, S. and {Sartoretti}, P. and {Soubiran}, C. and {Walton}, N.~A. and {Arenou}, F. and {Bailer-Jones}, C.~A.~L. and {Bastian}, U. and {Cropper}, M. and {Drimmel}, R. and {Katz}, D. and {Lattanzi}, M.~G. and {van Leeuwen}, F. and {Bakker}, J. and {Cacciari}, C. and {Casta{\~n}eda}, J. and {De Angeli}, F. and {Ducourant}, C. and {Fabricius}, C. and {Fouesneau}, M. and {Fr{\'e}mat}, Y. and {Guerra}, R. and {Guerrier}, A. and {Guiraud}, J. and {Jean-Antoine Piccolo}, A. and {Masana}, E. and {Messineo}, R. and {Mowlavi}, N. and {Nicolas}, C. and {Nienartowicz}, K. and {Pailler}, F. and {Panuzzo}, P. and {Riclet}, F. and {Roux}, W. and {Seabroke}, G.~M. and {Sordo}, R. and {Tanga}, P. and {Th{\'e}venin}, F. and {Gracia-Abril}, G. and {Portell}, J. and {Teyssier}, D. and {Altmann}, M. and {Andrae}, R. and {Bellas-Velidis}, I. and {Benson}, K. and {Berthier}, J. and {Blomme}, R. and {Brugaletta}, E. and {Burgess}, P.~W. and {Busso}, G. and {Carry}, B. and {Cellino}, A. and {Cheek}, N. and {Clementini}, G. and {Damerdji}, Y. and {Davidson}, M. and {Delchambre}, L. and {Dell'Oro}, A. and {Fern{\'a}ndez-Hern{\'a}ndez}, J. and {Galluccio}, L. and {Garc{\'\i}a-Lario}, P. and {Garcia-Reinaldos}, M. and {Gonz{\'a}lez-N{\'u}{\~n}ez}, J. and {Gosset}, E. and {Haigron}, R. and {Halbwachs}, J. -L. and {Hambly}, N.~C. and {Harrison}, D.~L. and {Hatzidimitriou}, D. and {Heiter}, U. and {Hern{\'a}ndez}, J. and {Hestroffer}, D. and {Hodgkin}, S.~T. and {Holl}, B. and {Jan{\ss}en}, K. and {Jevardat de Fombelle}, G. and {Jordan}, S. and {Krone-Martins}, A. and {Lanzafame}, A.~C. and {L{\"o}ffler}, W. and {Lorca}, A. and {Manteiga}, M. and {Marchal}, O. and {Marrese}, P.~M. and {Moitinho}, A. and {Mora}, A. and {Muinonen}, K. and {Osborne}, P. and {Pancino}, E. and {Pauwels}, T. and {Petit}, J. -M. and {Recio-Blanco}, A. and {Richards}, P.~J. and {Riello}, M. and {Rimoldini}, L. and {Robin}, A.~C. and {Roegiers}, T. and {Rybizki}, J. and {Sarro}, L.~M. and {Siopis}, C. and {Smith}, M. and {Sozzetti}, A. and {Ulla}, A. and {Utrilla}, E. and {van Leeuwen}, M. and {van Reeven}, W. and {Abbas}, U. and {Abreu Aramburu}, A. and {Accart}, S. and {Aerts}, C. and {Aguado}, J.~J. and {Ajaj}, M. and {Altavilla}, G. and {{\'A}lvarez}, M.~A. and {{\'A}lvarez Cid-Fuentes}, J. and {Alves}, J. and {Anderson}, R.~I. and {Anglada Varela}, E. and {Antoja}, T. and {Audard}, M. and {Baines}, D. and {Baker}, S.~G. and {Balaguer-N{\'u}{\~n}ez}, L. and {Balbinot}, E. and {Balog}, Z. and {Barache}, C. and {Barbato}, D. and {Barros}, M. and {Barstow}, M.~A. and {Bartolom{\'e}}, S. and {Bassilana}, J. -L. and {Bauchet}, N. and {Baudesson-Stella}, A. and {Becciani}, U. and {Bellazzini}, M. and {Bernet}, M. and {Bertone}, S. and {Bianchi}, L. and {Blanco-Cuaresma}, S. and {Boch}, T. and {Bombrun}, A. and {Bossini}, D. and {Bouquillon}, S. and {Bragaglia}, A. and {Bramante}, L. and {Breedt}, E. and {Bressan}, A. and {Brouillet}, N. and {Bucciarelli}, B. and {Burlacu}, A. and {Busonero}, D. and {Butkevich}, A.~G. and {Buzzi}, R. and {Caffau}, E. and {Cancelliere}, R. and {C{\'a}novas}, H. and {Cantat-Gaudin}, T. and {Carballo}, R. and {Carlucci}, T. and {Carnerero}, M.~I. and {Carrasco}, J.~M. and {Casamiquela}, L. and {Castellani}, M. and {Castro-Ginard}, A. and {Castro Sampol}, P. and {Chaoul}, L. and {Charlot}, P. and {Chemin}, L. and {Chiavassa}, A. and {Cioni}, M. -R.~L. and {Comoretto}, G. and {Cooper}, W.~J. and {Cornez}, T. and {Cowell}, S. and {Crifo}, F. and {Crosta}, M. and {Crowley}, C. and {Dafonte}, C. and {Dapergolas}, A. and {David}, M. and {David}, P.},
        title = "{Gaia Early Data Release 3. Summary of the contents and survey properties}",
      journal = {\aap},
         year = 2021,
        month = may,
       volume = {649},
          eid = {A1},
        pages = {A1},
          doi = {10.1051/0004-6361/202039657},
archivePrefix = {arXiv},
       eprint = {2012.01533},
 primaryClass = {astro-ph.GA},
       adsurl = {https://ui.adsabs.harvard.edu/abs/2021A&A...649A...1G}
}

@ARTICLE{Garcia2013,
       author = {{Garcia}, P.~J.~V. and {Benisty}, M. and {Dougados}, C. and {Bacciotti}, F. and {Clausse}, J. -M. and {Massi}, F. and {M{\'e}rand}, A. and {Petrov}, R. and {Weigelt}, G.},
        title = "{Pre-main-sequence binaries with tidally disrupted discs: the Br{\ensuremath{\gamma}} in HD 104237}",
      journal = {\mnras},
         year = 2013,
        month = apr,
       volume = {430},
       number = {3},
        pages = {1839-1853},
          doi = {10.1093/mnras/stt005},
archivePrefix = {arXiv},
       eprint = {1301.0276},
 primaryClass = {astro-ph.SR},
       adsurl = {https://ui.adsabs.harvard.edu/abs/2013MNRAS.430.1839G}
}

@ARTICLE{Grady2004,
       author = {{Grady}, C.~A. and {Woodgate}, B. and {Torres}, Carlos A.~O. and {Henning}, Th. and {Apai}, D. and {Rodmann}, J. and {Wang}, Hongchi and {Stecklum}, B. and {Linz}, H. and {Williger}, G.~M. and {Brown}, A. and {Wilkinson}, E. and {Harper}, G.~M. and {Herczeg}, G.~J. and {Danks}, A. and {Vieira}, G.~L. and {Malumuth}, E. and {Collins}, N.~R. and {Hill}, R.~S.},
        title = "{The Environment of the Optically Brightest Herbig Ae Star, HD 104237}",
      journal = {\apj},
         year = 2004,
        month = jun,
       volume = {608},
       number = {2},
        pages = {809-830},
          doi = {10.1086/420763},
       adsurl = {https://ui.adsabs.harvard.edu/abs/2004ApJ...608..809G}
}

@ARTICLE{Hales2014,
       author = {{Hales}, A.~S. and {De Gregorio-Monsalvo}, I. and {Montesinos}, B. and {Casassus}, S. and {Dent}, W.~F.~R. and {Dougados}, C. and {Eiroa}, C. and {Hughes}, A.~M. and {Garay}, G. and {Mardones}, D. and {M{\'e}nard}, F. and {Palau}, Aina and {P{\'e}rez}, S. and {Phillips}, N. and {Torrelles}, J.~M. and {Wilner}, D.},
        title = "{A CO Survey in Planet-forming Disks: Characterizing the Gas Content in the Epoch of Planet Formation}",
      journal = {\aj},
         year = 2014,
        month = sep,
       volume = {148},
       number = {3},
          eid = {47},
        pages = {47},
          doi = {10.1088/0004-6256/148/3/47},
archivePrefix = {arXiv},
       eprint = {1405.6966},
 primaryClass = {astro-ph.SR},
       adsurl = {https://ui.adsabs.harvard.edu/abs/2014AJ....148...47H}
}

@ARTICLE{Hartmann1998,
author = {{Hartmann}, L. and {Calvet}, N. and {Gullbring}, E. and {D'Alessio}, P.
},
title = "{Accretion and the Evolution of T Tauri Disks}",
journal = {\apj},
year = 1998,
month = mar,
volume = 495,
pages = {385-400},
doi = {10.1086/305277},
adsurl = {http://adsabs.harvard.edu/abs/1998ApJ...495..385H}
}

@ARTICLE{Heath2020,
       author = {{Heath}, R.~M. and {Nixon}, C.~J.},
        title = "{On the orbital evolution of binaries with circumbinary discs}",
      journal = {\aap},
         year = 2020,
        month = sep,
       volume = {641},
          eid = {A64},
        pages = {A64},
          doi = {10.1051/0004-6361/202038548},
archivePrefix = {arXiv},
       eprint = {2007.11592},
 primaryClass = {astro-ph.HE},
       adsurl = {https://ui.adsabs.harvard.edu/abs/2020A&A...641A..64H}
}

@ARTICLE{Hirsh2020,
       author = {{Hirsh}, Kieran and {Price}, Daniel J. and {Gonzalez}, Jean-Fran{\c{c}}ois and {Ubeira-Gabellini}, M. Giulia and {Ragusa}, Enrico},
        title = "{On the cavity size in circumbinary discs}",
      journal = {\mnras},
         year = 2020,
        month = oct,
       volume = {498},
       number = {2},
        pages = {2936-2947},
          doi = {10.1093/mnras/staa2536},
archivePrefix = {arXiv},
       eprint = {2008.08008},
 primaryClass = {astro-ph.EP},
       adsurl = {https://ui.adsabs.harvard.edu/abs/2020MNRAS.498.2936H}
}

@ARTICLE{Huhn2025,
       author = {{H{\"u}hn}, L.-A. and {Jiang}, H.-C. and {Dullemond}, C.~P.},
        title = "{Late accretion offers pathway to misaligned disk around the planet-hosting IRAS 04125+2902}",
      journal = {\aap},
         year = 2025,
        month = sep,
       volume = {701},
          eid = {L15},
        pages = {L15},
          doi = {10.1051/0004-6361/202555391},
archivePrefix = {arXiv},
       eprint = {2509.06564},
 primaryClass = {astro-ph.EP},
       adsurl = {https://ui.adsabs.harvard.edu/abs/2025A&A...701L..15H}
}

@ARTICLE{Iglesias2023,
       author = {{Iglesias}, Daniela P. and {Pani{\'c}}, Olja and {van den Ancker}, Mario and {Petr-Gotzens}, Monika G. and {Siess}, Lionel and {Vioque}, Miguel and {Pascucci}, Ilaria and {Oudmaijer}, Ren{\'e} and {Miley}, James},
        title = "{X-shooter survey of young intermediate-mass stars - I. Stellar characterization and disc evolution}",
      journal = {\mnras},
         year = 2023,
        month = mar,
       volume = {519},
       number = {3},
        pages = {3958-3975},
          doi = {10.1093/mnras/stac3619},
archivePrefix = {arXiv},
       eprint = {2212.06791},
 primaryClass = {astro-ph.SR},
       adsurl = {https://ui.adsabs.harvard.edu/abs/2023MNRAS.519.3958I}
}

@ARTICLE{Jankovic2021,
       author = {{Jankovic}, Marija R. and {Owen}, James E. and {Mohanty}, Subhanjoy and {Tan}, Jonathan C.},
        title = "{MRI-active inner regions of protoplanetary discs. I. A detailed model of disc structure}",
      journal = {\mnras},
         year = 2021,
        month = jun,
       volume = {504},
       number = {1},
        pages = {280-299},
          doi = {10.1093/mnras/stab920},
archivePrefix = {arXiv},
       eprint = {2102.12831},
 primaryClass = {astro-ph.EP},
       adsurl = {https://ui.adsabs.harvard.edu/abs/2021MNRAS.504..280J}
}

@ARTICLE{Jankovic2022,
       author = {{Jankovic}, Marija R. and {Mohanty}, Subhanjoy and {Owen}, James E. and {Tan}, Jonathan C.},
        title = "{MRI-active inner regions of protoplanetary discs - II. Dependence on dust, disc, and stellar parameters}",
      journal = {\mnras},
         year = 2022,
        month = feb,
       volume = {509},
       number = {4},
        pages = {5974-5991},
          doi = {10.1093/mnras/stab3370},
archivePrefix = {arXiv},
       eprint = {2108.12332},
 primaryClass = {astro-ph.SR},
       adsurl = {https://ui.adsabs.harvard.edu/abs/2022MNRAS.509.5974J}
}

@ARTICLE{Juhasz2025,
       author = {{Juh{\'a}sz}, T{\'\i}mea and {Varga}, J{\'o}zsef and {{\'A}brah{\'a}m}, P{\'e}ter and {K{\'o}sp{\'a}l}, {\'A}gnes and {Lykou}, Foteini and {Chen}, Lei and {Mo{\'o}r}, Attila and {Cruz-S{\'a}enz de Miera}, Fernando and {Lopez}, Bruno and {Matter}, Alexis and {van Boekel}, Roy and {Hogerheijde}, Michiel and {Abello}, Margaux and {Augereau}, Jean-Charles and {Boley}, Paul and {Danchi}, William C. and {Henning}, Thomas and {Letessier}, Mathis and {Ma}, Jie and {Priolet}, Philippe and {Scheuck}, Marten and {Weigelt}, Gerd and {Wolf}, Sebastian},
        title = "{Evidence for an Accretion Bridge in the DX Cha Circumbinary System from VLTI/MATISSE Observations}",
      journal = {\apj},
         year = 2025,
        month = mar,
       volume = {982},
       number = {1},
          eid = {36},
        pages = {36},
          doi = {10.3847/1538-4357/adb727},
archivePrefix = {arXiv},
       eprint = {2502.11722},
 primaryClass = {astro-ph.SR},
       adsurl = {https://ui.adsabs.harvard.edu/abs/2025ApJ...982...36J}
}

@ARTICLE{Kennedy2012,
author = {{Kennedy}, G.~M. and {Wyatt}, M.~C. and {Sibthorpe}, B. and 
{Duch{\^e}ne}, G. and {Kalas}, P. and {Matthews}, B.~C. and 
{Greaves}, J.~S. and {Su}, K.~Y.~L. and {Fitzgerald}, M.~P.},
title = "{99 Herculis: host to a circumbinary polar-ring debris disc}",
journal = {\mnras},
archivePrefix = "arXiv",
eprint = {1201.1911},
primaryClass = "astro-ph.EP",
year = 2012,
month = apr,
volume = 421,
pages = {2264-2276},
doi = {10.1111/j.1365-2966.2012.20448.x},
adsurl = {http://adsabs.harvard.edu/abs/2012MNRAS.421.2264K}
}

@ARTICLE{Kennedy2019,
       author = {{Kennedy}, Grant M. and {Matr{\`a}}, Luca and {Facchini}, Stefano and
         {Milli}, Julien and {Pani{\'c}}, Olja and {Price}, Daniel and
         {Wilner}, David J. and {Wyatt}, Mark C. and {Yelverton}, Ben M.},
        title = "{Publisher Correction: A circumbinary protoplanetary disk in a polar configuration}",
      journal = {Nature Astronomy},
         year = "2019",
        month = "Feb",
       volume = {3},
        pages = {278-278},
          doi = {10.1038/s41550-019-0715-1},
archivePrefix = {arXiv},
       eprint = {1901.05018},
 primaryClass = {astro-ph.EP},
       adsurl = {https://ui.adsabs.harvard.edu/abs/2019NatAs...3..278K}
}

@ARTICLE{Korycansky1995,
       author = {{Korycansky}, D.~G. and {Pringle}, J.~E.},
        title = "{Axisymmetric waves in polytropic accretion discs}",
      journal = {\mnras},
         year = 1995,
        month = feb,
       volume = {272},
       number = {3},
        pages = {618-624},
          doi = {10.1093/mnras/272.3.618},
       adsurl = {https://ui.adsabs.harvard.edu/abs/1995MNRAS.272..618K}
}

@ARTICLE{Kostov2014,
author = {{Kostov}, V.~B. and {McCullough}, P.~R. and {Carter}, J.~A. and 
{Deleuil}, M. and {D{\'{\i}}az}, R.~F. and {Fabrycky}, D.~C. and 
{H{\'e}brard}, G. and {Hinse}, T.~C. and {et al.}},
title = "{Kepler-413b: A Slightly Misaligned, Neptune-size Transiting Circumbinary Planet}",
journal = {\apj},
archivePrefix = "arXiv",
eprint = {1401.7275},
primaryClass = "astro-ph.EP",
year = 2014,
month = mar,
volume = 784,
eid = {14},
pages = {14},
doi = {10.1088/0004-637X/784/1/14},
adsurl = {http://adsabs.harvard.edu/abs/2014ApJ...784...14K}
}

@ARTICLE{Kostov2016,
author = {{Kostov}, V.~B. and {Orosz}, J.~A. and {Welsh}, W.~F. and {Doyle}, L.~R. and 
{Fabrycky}, D.~C. and {Haghighipour}, N. and {Quarles}, B. and 
{Short}, D.~R. and {et al.}
},
title = "{Kepler-1647b: The Largest and Longest-period Kepler Transiting Circumbinary Planet}",
journal = {\apj},
archivePrefix = "arXiv",
eprint = {1512.00189},
primaryClass = "astro-ph.EP",
year = 2016,
month = aug,
volume = 827,
eid = {86},
pages = {86},
doi = {10.3847/0004-637X/827/1/86},
adsurl = {http://adsabs.harvard.edu/abs/2016ApJ...827...86K}
}

@ARTICLE{Kostov2020,
       author = {{Kostov}, Veselin B. and {Orosz}, Jerome A. and {Feinstein}, Adina D. and {Welsh}, William F. and {Cukier}, Wolf and {Haghighipour}, Nader and {Quarles}, Billy and {Martin}, David V. and {Montet}, Benjamin T. and {Torres}, Guillermo and {Triaud}, Amaury H.~M.~J. and {Barclay}, Thomas and {Boyd}, Patricia and {Briceno}, Cesar and {Cameron}, Andrew Collier and {Correia}, Alexandre C.~M. and {Gilbert}, Emily A. and {Gill}, Samuel and {Gillon}, Micha{\"e}l and {Haqq-Misra}, Jacob and {Hellier}, Coel and {Dressing}, Courtney and {Fabrycky}, Daniel C. and {Furesz}, Gabor and {Jenkins}, Jon M. and {Kane}, Stephen R. and {Kopparapu}, Ravi and {Hod{\v{z}}i{\'c}}, Vedad Kunovac and {Latham}, David W. and {Law}, Nicholas and {Levine}, Alan M. and {Li}, Gongjie and {Lintott}, Chris and {Lissauer}, Jack J. and {Mann}, Andrew W. and {Mazeh}, Tsevi and {Mardling}, Rosemary and {Maxted}, Pierre F.~L. and {Eisner}, Nora and {Pepe}, Francesco and {Pepper}, Joshua and {Pollacco}, Don and {Quinn}, Samuel N. and {Quintana}, Elisa V. and {Rowe}, Jason F. and {Ricker}, George and {Rose}, Mark E. and {Seager}, S. and {Santerne}, Alexandre and {S{\'e}gransan}, Damien and {Short}, Donald R. and {Smith}, Jeffrey C. and {Standing}, Matthew R. and {Tokovinin}, Andrei and {Trifonov}, Trifon and {Turner}, Oliver and {Twicken}, Joseph D. and {Udry}, St{\'e}phane and {Vanderspek}, Roland and {Winn}, Joshua N. and {Wolf}, Eric T. and {Ziegler}, Carl and {Ansorge}, Peter and {Barnet}, Frank and {Bergeron}, Joel and {Huten}, Marc and {Pappa}, Giuseppe and {van der Straeten}, Timo},
        title = "{TOI-1338: TESS' First Transiting Circumbinary Planet}",
      journal = {\aj},
         year = 2020,
        month = jun,
       volume = {159},
       number = {6},
          eid = {253},
        pages = {253},
          doi = {10.3847/1538-3881/ab8a48},
archivePrefix = {arXiv},
       eprint = {2004.07783},
 primaryClass = {astro-ph.EP},
       adsurl = {https://ui.adsabs.harvard.edu/abs/2020AJ....159..253K}
}

@ARTICLE{Kuffmeier2021,
       author = {{Kuffmeier}, M. and {Dullemond}, C.~P. and {Reissl}, S. and {Goicovic}, F.~G.},
        title = "{Misaligned disks induced by infall}",
      journal = {\aap},
         year = 2021,
        month = dec,
       volume = {656},
          eid = {A161},
        pages = {A161},
          doi = {10.1051/0004-6361/202039614},
archivePrefix = {arXiv},
       eprint = {2110.04309},
 primaryClass = {astro-ph.SR},
       adsurl = {https://ui.adsabs.harvard.edu/abs/2021A&A...656A.161K}
}

@ARTICLE{Lazareff2017,
       author = {{Lazareff}, B. and {Berger}, J. -P. and {Kluska}, J. and {Le Bouquin}, J. -B. and {Benisty}, M. and {Malbet}, F. and {Koen}, C. and {Pinte}, C. and {Thi}, W. -F. and {Absil}, O. and {Baron}, F. and {Delboulb{\'e}}, A. and {Duvert}, G. and {Isella}, A. and {Jocou}, L. and {Juhasz}, A. and {Kraus}, S. and {Lachaume}, R. and {M{\'e}nard}, F. and {Millan-Gabet}, R. and {Monnier}, J.~D. and {Moulin}, T. and {Perraut}, K. and {Rochat}, S. and {Soulez}, F. and {Tallon}, M. and {Thi{\'e}baut}, E. and {Traub}, W. and {Zins}, G.},
        title = "{Structure of Herbig AeBe disks at the milliarcsecond scale . A statistical survey in the H band using PIONIER-VLTI}",
      journal = {\aap},
         year = 2017,
        month = mar,
       volume = {599},
          eid = {A85},
        pages = {A85},
          doi = {10.1051/0004-6361/201629305},
archivePrefix = {arXiv},
       eprint = {1611.08428},
 primaryClass = {astro-ph.SR},
       adsurl = {https://ui.adsabs.harvard.edu/abs/2017A&A...599A..85L}
}

@ARTICLE{Leinert2003,
       author = {{Leinert}, Ch. and {Graser}, U. and {Przygodda}, F. and {Waters}, L.~B.~F.~M. and {Perrin}, G. and {Jaffe}, W. and {Lopez}, B. and {Bakker}, E.~J. and {B{\"o}hm}, A. and {Chesneau}, O. and {Cotton}, W.~D. and {Damstra}, S. and {de Jong}, J. and {Glazenborg-Kluttig}, A.~W. and {Grimm}, B. and {Hanenburg}, H. and {Laun}, W. and {Lenzen}, R. and {Ligori}, S. and {Mathar}, R.~J. and {Meisner}, J. and {Morel}, S. and {Morr}, W. and {Neumann}, U. and {Pel}, J. -W. and {Schuller}, P. and {Rohloff}, R. -R. and {Stecklum}, B. and {Storz}, C. and {von der L{\"u}he}, O. and {Wagner}, K.},
        title = "{MIDI {\textendash} the 10 {\ensuremath{\mu}}m instrument on the VLTI}",
      journal = {\apss},
         year = 2003,
        month = aug,
       volume = {286},
       number = {1},
        pages = {73-83},
          doi = {10.1023/A:1026158127732},
       adsurl = {https://ui.adsabs.harvard.edu/abs/2003Ap&SS.286...73L}
}

@ARTICLE{Leung2013,
       author = {{Leung}, Gene C.~K. and {Lee}, Man Hoi},
        title = "{An Analytic Theory for the Orbits of Circumbinary Planets}",
      journal = {\apj},
         year = 2013,
        month = feb,
       volume = {763},
       number = {2},
          eid = {107},
        pages = {107},
          doi = {10.1088/0004-637X/763/2/107},
archivePrefix = {arXiv},
       eprint = {1212.2545},
 primaryClass = {astro-ph.EP},
       adsurl = {https://ui.adsabs.harvard.edu/abs/2013ApJ...763..107L}
}

@ARTICLE{Li2016,
author = {{Li}, G. and {Holman}, M.~J. and {Tao}, M.},
title = "{Uncovering Circumbinary Planetary Architectural Properties from Selection Biases}",
journal = {\apj},
archivePrefix = "arXiv",
eprint = {1608.01768},
primaryClass = "astro-ph.EP",
year = 2016,
month = nov,
volume = 831,
eid = {96},
pages = {96},
doi = {10.3847/0004-637X/831/1/96},
adsurl = {http://adsabs.harvard.edu/abs/2016ApJ...831...96L}
}

@ARTICLE{LP1993,
author = {{Lubow}, S.~H. and {Pringle}, J.~E.},
title = "{Wave propagation in accretion disks - Axisymmetric case}",
journal = {ApJ},
year = 1993,
month = may,
volume = 409,
pages = {360-371},
doi = {10.1086/172669},
adsurl = {http://adsabs.harvard.edu/abs/1993ApJ...409..360L}
}

@ARTICLE{Lubow1998,
       author = {{Lubow}, S.~H. and {Ogilvie}, G.~I.},
        title = "{Three-dimensional Waves Generated at Lindblad Resonances in Thermally Stratified Disks}",
      journal = {\apj},
         year = 1998,
        month = sep,
       volume = {504},
       number = {2},
        pages = {983-995},
          doi = {10.1086/306104},
archivePrefix = {arXiv},
       eprint = {astro-ph/9804063},
 primaryClass = {astro-ph},
       adsurl = {https://ui.adsabs.harvard.edu/abs/1998ApJ...504..983L}
}

@ARTICLE{LP1974,
author = {{Lynden-Bell}, D. and {Pringle}, J.~E.},
title = "{The evolution of viscous discs and the origin of the nebular variables.}",
journal = {MNRAS},
year = 1974,
month = sep,
volume = 168,
pages = {603-637},
adsurl = {http://adsabs.harvard.edu/abs/1974MNRAS.168..603L}
}

@ARTICLE{Lynden-Bell1972,
       author = {{Lynden-Bell}, D. and {Kalnajs}, A.~J.},
        title = "{On the generating mechanism of spiral structure}",
      journal = {\mnras},
         year = 1972,
        month = jan,
       volume = {157},
        pages = {1},
          doi = {10.1093/mnras/157.1.1},
       adsurl = {https://ui.adsabs.harvard.edu/abs/1972MNRAS.157....1L}
}

@ARTICLE{McKee2007,
author = {{McKee}, C.~F. and {Ostriker}, E.~C.},
title = "{Theory of Star Formation}",
journal = {\araa},
archivePrefix = "arXiv",
eprint = {0707.3514},
year = 2007,
month = sep,
volume = 45,
pages = {565-687},
doi = {10.1146/annurev.astro.45.051806.110602},
adsurl = {http://adsabs.harvard.edu/abs/2007ARA%26A..45..565M}
}

@ARTICLE{Martin2019,
       author = {{Martin}, R.~G. and {Nixon}, C.~J. and {Pringle}, J.~E. and {Livio}, M.},
        title = "{On the physical nature of accretion disc viscosity}",
      journal = {\na},
         year = 2019,
        month = jul,
       volume = {70},
        pages = {7-11},
          doi = {10.1016/j.newast.2019.01.001},
archivePrefix = {arXiv},
       eprint = {1901.01580},
 primaryClass = {astro-ph.HE},
       adsurl = {https://ui.adsabs.harvard.edu/abs/2019NewA...70....7M}
}

@ARTICLE{Mendigutia2011b,
       author = {{Mendigut{\'\i}a}, I. and {Calvet}, N. and {Montesinos}, B. and {Mora}, A. and {Muzerolle}, J. and {Eiroa}, C. and {Oudmaijer}, R.~D. and {Mer{\'\i}n}, B.},
        title = "{Accretion rates and accretion tracers of Herbig Ae/Be stars}",
      journal = {\aap},
         year = 2011,
        month = nov,
       volume = {535},
          eid = {A99},
        pages = {A99},
          doi = {10.1051/0004-6361/201117444},
archivePrefix = {arXiv},
       eprint = {1109.3288},
 primaryClass = {astro-ph.SR},
       adsurl = {https://ui.adsabs.harvard.edu/abs/2011A&A...535A..99M}
}

@article{Nidhi2023,
    author = {Nidhi, S and Mathew, Blesson and Shridharan, B and Arun, R and Anusha, R and Kartha, Sreeja S},
    title = {Spectroscopic study of Herbig Ae/Be stars in the Galactic anti-centre region from LAMOST DR5},
    journal = {Monthly Notices of the Royal Astronomical Society},
    volume = {524},
    number = {4},
    pages = {5166-5181},
    year = {2023},
    month = {07},
    issn = {0035-8711},
    doi = {10.1093/mnras/stad2067},
    url = {https://doi.org/10.1093/mnras/stad2067},
    eprint = {https://academic.oup.com/mnras/article-pdf/524/4/5166/51016993/stad2067.pdf},
}

@ARTICLE{Nixonetal2011a,
author = {{Nixon}, C.~J. and {Cossins}, P.~J. and {King}, A.~R. and {Pringle}, J.~E.
},
title = "{Retrograde accretion and merging supermassive black holes}",
journal = {MNRAS},
archivePrefix = "arXiv",
eprint = {1011.1914},
primaryClass = "astro-ph.HE",
year = 2011,
month = apr,
volume = 412,
pages = {1591-1598},
doi = {10.1111/j.1365-2966.2010.17952.x},
adsurl = {http://adsabs.harvard.edu/abs/2011MNRAS.412.1591N}
}

@ARTICLE{Nixonetal2011b,
author = {{Nixon}, C.~J. and {King}, A.~R. and {Pringle}, J.~E.},
title = "{The final parsec problem: aligning a binary with an external accretion disc}",
journal = {MNRAS},
archivePrefix = "arXiv",
eprint = {1107.5056},
primaryClass = "astro-ph.GA",
year = 2011,
month = oct,
volume = 417,
pages = {L66-L69},
doi = {10.1111/j.1745-3933.2011.01121.x},
adsurl = {http://adsabs.harvard.edu/abs/2011MNRAS.417L..66N}
}

@ARTICLE{Nixon2012,
author = {{Nixon}, C.~J.},
title = "{Stable counteralignment of a circumbinary disc}",
journal = {MNRAS},
archivePrefix = "arXiv",
eprint = {1204.4185},
primaryClass = "astro-ph.HE",
year = 2012,
month = jul,
volume = 423,
pages = {2597-2600},
doi = {10.1111/j.1365-2966.2012.21072.x},
adsurl = {http://adsabs.harvard.edu/abs/2012MNRAS.423.2597N}
}

@ARTICLE{Nixon2015,
author = {{Nixon}, C. and {Lubow}, S.~H.},
title = "{Resonances in retrograde circumbinary discs}",
journal = {MNRAS},
archivePrefix = "arXiv",
eprint = {1501.07277},
primaryClass = "astro-ph.HE",
year = 2015,
month = apr,
volume = 448,
pages = {3472-3483},
doi = {10.1093/mnras/stv166},
adsurl = {http://adsabs.harvard.edu/abs/2015MNRAS.448.3472N}
}

@ARTICLE{Orosz2012a,
       author = {{Orosz}, Jerome A. and {Welsh}, William F. and {Carter}, Joshua A. and {Brugamyer}, Erik and {Buchhave}, Lars A. and {Cochran}, William D. and {Endl}, Michael and {Ford}, Eric B. and {MacQueen}, Phillip and {Short}, Donald R. and {Torres}, Guillermo and {Windmiller}, Gur and {Agol}, Eric and {Barclay}, Thomas and {Caldwell}, Douglas A. and {Clarke}, Bruce D. and {Doyle}, Laurance R. and {Fabrycky}, Daniel C. and {Geary}, John C. and {Haghighipour}, Nader and {Holman}, Matthew J. and {Ibrahim}, Khadeejah A. and {Jenkins}, Jon M. and {Kinemuchi}, Karen and {Li}, Jie and {Lissauer}, Jack J. and {Pr{\v{s}}a}, Andrej and {Ragozzine}, Darin and {Shporer}, Avi and {Still}, Martin and {Wade}, Richard A.},
        title = "{The Neptune-sized Circumbinary Planet Kepler-38b}",
      journal = {\apj},
         year = 2012,
        month = oct,
       volume = {758},
       number = {2},
          eid = {87},
        pages = {87},
          doi = {10.1088/0004-637X/758/2/87},
archivePrefix = {arXiv},
       eprint = {1208.3712},
 primaryClass = {astro-ph.SR},
       adsurl = {https://ui.adsabs.harvard.edu/abs/2012ApJ...758...87O}
}

@ARTICLE{Orosz2012b,
       author = {{Orosz}, Jerome A. and {Welsh}, William F. and {Carter}, Joshua A. and {Fabrycky}, Daniel C. and {Cochran}, William D. and {Endl}, Michael and {Ford}, Eric B. and {Haghighipour}, Nader and {MacQueen}, Phillip J. and {Mazeh}, Tsevi and {Sanchis-Ojeda}, Roberto and {Short}, Donald R. and {Torres}, Guillermo and {Agol}, Eric and {Buchhave}, Lars A. and {Doyle}, Laurance R. and {Isaacson}, Howard and {Lissauer}, Jack J. and {Marcy}, Geoffrey W. and {Shporer}, Avi and {Windmiller}, Gur and {Barclay}, Thomas and {Boss}, Alan P. and {Clarke}, Bruce D. and {Fortney}, Jonathan and {Geary}, John C. and {Holman}, Matthew J. and {Huber}, Daniel and {Jenkins}, Jon M. and {Kinemuchi}, Karen and {Kruse}, Ethan and {Ragozzine}, Darin and {Sasselov}, Dimitar and {Still}, Martin and {Tenenbaum}, Peter and {Uddin}, Kamal and {Winn}, Joshua N. and {Koch}, David G. and {Borucki}, William J.},
        title = "{Kepler-47: A Transiting Circumbinary Multiplanet System}",
      journal = {Science},
         year = 2012,
        month = sep,
       volume = {337},
       number = {6101},
        pages = {1511},
          doi = {10.1126/science.1228380},
archivePrefix = {arXiv},
       eprint = {1208.5489},
 primaryClass = {astro-ph.SR},
       adsurl = {https://ui.adsabs.harvard.edu/abs/2012Sci...337.1511O}
}

@ARTICLE{Pelkonen2025,
       author = {{Pelkonen}, V.-M. and {Padoan}, P. and {Juvela}, M. and {Haugb{\o}lle}, T. and {Nordlund}, {\r{A}}.},
        title = "{Origin and evolution of angular momentum of class II disks}",
      journal = {\aap},
         year = 2025,
        month = feb,
       volume = {694},
          eid = {A327},
        pages = {A327},
          doi = {10.1051/0004-6361/202450682},
archivePrefix = {arXiv},
       eprint = {2405.06520},
 primaryClass = {astro-ph.SR},
       adsurl = {https://ui.adsabs.harvard.edu/abs/2025A&A...694A.327P}
}

@ARTICLE{Penzlin2024,
       author = {{Penzlin}, Anna B.~T. and {Booth}, Richard A. and {Nelson}, Richard P. and {Sch{\"a}fer}, Christoph M. and {Kley}, Wilhelm},
        title = "{Viscous circumbinary protoplanetary discs - I. Structure of the inner cavity}",
      journal = {\mnras},
         year = 2024,
        month = aug,
       volume = {532},
       number = {3},
        pages = {3166-3179},
          doi = {10.1093/mnras/stae1689},
archivePrefix = {arXiv},
       eprint = {2407.07243},
 primaryClass = {astro-ph.EP},
       adsurl = {https://ui.adsabs.harvard.edu/abs/2024MNRAS.532.3166P}
}

@ARTICLE{Pierens2023,
       author = {{Pierens}, Arnaud and {Nelson}, Richard P.},
        title = "{Three-dimensional evolution of radiative circumbinary discs: The size and shape of the inner cavity}",
      journal = {\aap},
         year = 2023,
        month = feb,
       volume = {670},
          eid = {A112},
        pages = {A112},
          doi = {10.1051/0004-6361/202244828},
archivePrefix = {arXiv},
       eprint = {2211.03816},
 primaryClass = {astro-ph.EP},
       adsurl = {https://ui.adsabs.harvard.edu/abs/2023A&A...670A.112P}
}

@ARTICLE{Price2018,
       author = {{Price}, Daniel J. and {Wurster}, James and {Tricco}, Terrence S. and {Nixon}, Chris and {Toupin}, St{\'e}ven and {Pettitt}, Alex and {Chan}, Conrad and {Mentiplay}, Daniel and {Laibe}, Guillaume and {Glover}, Simon and {Dobbs}, Clare and {Nealon}, Rebecca and {Liptai}, David and {Worpel}, Hauke and {Bonnerot}, Cl{\'e}ment and {Dipierro}, Giovanni and {Ballabio}, Giulia and {Ragusa}, Enrico and {Federrath}, Christoph and {Iaconi}, Roberto and {Reichardt}, Thomas and {Forgan}, Duncan and {Hutchison}, Mark and {Constantino}, Thomas and {Ayliffe}, Ben and {Hirsh}, Kieran and {Lodato}, Giuseppe},
        title = "{Phantom: A Smoothed Particle Hydrodynamics and Magnetohydrodynamics Code for Astrophysics}",
      journal = {\pasa},
         year = 2018,
        month = sep,
       volume = {35},
          eid = {e031},
        pages = {e031},
          doi = {10.1017/pasa.2018.25},
archivePrefix = {arXiv},
       eprint = {1702.03930},
 primaryClass = {astro-ph.IM},
       adsurl = {https://ui.adsabs.harvard.edu/abs/2018PASA...35...31P}
}

@ARTICLE{Pringle1981,
author = {{Pringle}, J.~E.},
title = "{Accretion discs in astrophysics}",
journal = {ARA\&A},
year = 1981,
volume = 19,
pages = {137-162},
doi = {10.1146/annurev.aa.19.090181.001033},
adsurl = {http://adsabs.harvard.edu/abs/1981ARA%26A..19..137P}
}

@ARTICLE{Pringle1991,
author = {{Pringle}, J.~E.},
title = "{The properties of external accretion discs}",
journal = {MNRAS},
year = 1991,
month = feb,
volume = 248,
pages = {754-759},
adsurl = {http://adsabs.harvard.edu/abs/1991MNRAS.248..754P}
}

@ARTICLE{Roedig2014,
       author = {{Roedig}, Constanze and {Sesana}, Alberto},
        title = "{Migration of massive black hole binaries in self-gravitating discs: retrograde versus prograde}",
      journal = {\mnras},
         year = 2014,
        month = apr,
       volume = {439},
       number = {4},
        pages = {3476-3489},
          doi = {10.1093/mnras/stu194},
archivePrefix = {arXiv},
       eprint = {1307.6283},
 primaryClass = {astro-ph.HE},
       adsurl = {https://ui.adsabs.harvard.edu/abs/2014MNRAS.439.3476R}
}

@ARTICLE{Rosotti:2023,
       author = {{Rosotti}, Giovanni P.},
        title = "{Empirical constraints on turbulence in proto-planetary discs}",
      journal = {\nar},
         year = 2023,
        month = jun,
       volume = {96},
          eid = {101674},
        pages = {101674},
          doi = {10.1016/j.newar.2023.101674},
archivePrefix = {arXiv},
       eprint = {2302.01433},
 primaryClass = {astro-ph.EP},
       adsurl = {https://ui.adsabs.harvard.edu/abs/2023NewAR..9601674R}
}
\bibliographystyle{aasjournal}

\end{document}